\documentclass[letterpaper,twocolumn,10pt]{article}

\usepackage{usenix,epsfig,endnotes}
\usepackage[numbers,sort&compress]{natbib}
\usepackage{multirow}
\usepackage{subfig}
\usepackage{appendix}
\usepackage{graphicx}
\usepackage{grffile}

\newcounter{subcopyrightbox@save}
\usepackage[font=bf]{caption}
\usepackage{color, url}
\usepackage{xspace} 
\usepackage{mathrsfs}
\usepackage{amssymb}
\usepackage{amsmath}
\usepackage{epstopdf}
\usepackage{multicol}
\usepackage{balance}
\usepackage{algorithm}
\usepackage{algorithmic}

\usepackage{bbold}

\newtheorem{theorem}{Theorem}

\newtheorem{definition}[theorem]{Definition}
\newcommand{\argmax}{\operatornamewithlimits{argmax}}
\newcommand{\argmin}{\operatornamewithlimits{argmin}}
\newcommand{\myparatight}[1]{\smallskip\noindent{\bf {#1}:}~}

\newenvironment{packeditemize}{\begin{list}{$\bullet$}{\setlength{\itemsep}{0.2pt}\addtolength{\labelwidth}{-4pt}\setlength{\leftmargin}{\labelwidth}\setlength{\listparindent}{\parindent}\setlength{\parsep}{1pt}\setlength{\topsep}{0pt}}}{\end{list}}

\graphicspath{ {../fig/} }

\usepackage{fancyhdr}
\begin{document}
%
\title{AttriGuard: A Practical Defense Against Attribute Inference Attacks via Adversarial Machine Learning}

\author{
{\rm Jinyuan Jia} \\
Duke University\\
jinyuan.jia@duke.edu
\and
{\rm Neil Zhenqiang Gong} \\
Duke University\\
neil.gong@duke.edu
}
\maketitle

\begin{abstract}
Users in various web and mobile applications are vulnerable to \emph{attribute inference attacks}, in which an attacker leverages a machine learning classifier to infer a target user's private attributes (e.g., location, sexual orientation, political view) from its public data (e.g., rating scores, page likes). 
Existing defenses leverage game theory or heuristics based on correlations between the public data and attributes. These defenses are not practical. 
Specifically, game-theoretic defenses require solving intractable optimization problems,  
while correlation-based defenses incur large utility loss of users' public data. 

In this paper, we present \emph{AttriGuard}, a practical defense against attribute inference attacks. AttriGuard is computationally tractable and has small utility loss.  Our AttriGuard works in two phases. Suppose we aim to protect a user's private attribute. In Phase I, for each value of the attribute, we find a minimum noise such that if we add the noise to the user's public data, then the attacker's classifier is very likely to infer the attribute value for the user. We find the minimum noise via adapting existing \emph{evasion attacks} in adversarial machine learning. In Phase II, we sample one attribute value according to a certain probability distribution and add the corresponding noise found in Phase I to the user's public data. We formulate finding the probability distribution as solving a constrained convex optimization problem. We extensively evaluate AttriGuard and compare it with existing methods using a real-world dataset. Our results show that AttriGuard substantially outperforms existing methods.  Our work is the first one that shows evasion attacks can be used as defensive techniques for privacy protection. 

\end{abstract}
\section{Introduction}



\emph{Attribute inference attacks} are emerging threats to user privacy in various application domains ranging from social media~\cite{Zheleva09,Chaabane12,kosinski2013private,Gong14,GongAttriInferSEC16,AttriInfer,Gong18tops} to recommender systems~\cite{Otterbacher10,weinsberg2012blurme} to mobile platforms~\cite{MichalevskyPowerSEC,Narain16Oakland}. 
In an attribute inference attack, an attacker aims to infer a user's {private attributes} (e.g., location, gender, sexual orientation, and/or political view) via leveraging its public data. 
For instance, in social media, a user's public data could be the list of pages that the user liked on Facebook. Given these page likes, an attacker can use a machine learning classifier to accurately infer the user's various private attributes including, but not limited to, gender, sexual orientation, and political view~\cite{kosinski2013private}. Such inferred attributes can be further leveraged to deliver personalized advertisements to users~\cite{Cambridge}. 
 In recommender systems, a user's public data could be the list of items (e.g., movies, mobile apps, videos) that the user rated. Given the rating scores, an attacker can use a classifier to infer a user's gender with an alarming accuracy~\cite{weinsberg2012blurme}.  Attribute inference attacks can successfully infer a user's private attributes via its public data because users' private attributes are statistically correlated with their public data.


We represent a user's public data as a \emph{vector}. For instance, in recommender systems, an entry of the vector is the rating score the user gave to the corresponding item or 0 if the user did not rate the item. A defense against attribute inference attacks essentially adds noise to a user's public data vector (i.e., modify certain entries of the vector) with a goal to decrease the inference accuracy of an attacker. 
One category of defenses (e.g.,~\cite{ShokriCCS12,ShokriPETS15,privacygamelocation16,Calmon:2012}) against general inference attacks leverage game theory. In these methods, an attacker performs the optimal inference attack based on the knowledge of the defense, while the defender defends against the optimal inference attack. These game-theoretic methods have theoretical privacy guarantees, i.e., they  defend against the optimal inference attack. However, they are computationally intractable when applied to attribute inference attacks. For instance, in Appendix~\ref{gametheory}, we extend the game-theoretic method from Shokri et al.~\cite{ShokriCCS12}  
to attribute inference attacks. 
The computation cost to solve the formulated optimization problem is \emph{exponential} to the dimensionality of the public data vector and the public data vector often has high dimensionality in practice. 


To address the computational challenges, several studies~\cite{weinsberg2012blurme,friendbasedDefense,ChenObfuscationPETS14,Salamatian:2015} proposed to trade theoretical privacy guarantees for computational tractability. Specifically, Salamatian et al.~\cite{Salamatian:2015} proposed to quantize the public data to approximately solve the game-theoretic optimization problem~\cite{Calmon:2012}. Several other methods~\cite{weinsberg2012blurme,friendbasedDefense,ChenObfuscationPETS14} leverage correlation-based heuristics, e.g., they modify the public data entries that have large correlations with the private attribute values that do not belong to a user. However, these methods suffer from one or two key limitations. First, as we will demonstrate in our experiments, they incur large utility loss, i.e., they add a large amount of noise to a user's public data. Second, some of them~\cite{weinsberg2012blurme,friendbasedDefense,ChenObfuscationPETS14} require the defender to have direct access to a user's private attribute value, in order to compute the correlations between public data entries and private attribute values that do not belong to the user.  
Such requirement introduces usability issue and additional privacy concerns. Specifically, a user needs to specify its attribute value to the defender, which makes it inconvenient for users. Moreover, the defender becomes a single point of failure, i.e., when the defender is compromised, the private attribute values of all users are compromised. 

To summarize, existing defense methods against attribute inference attacks are not practical. Specifically, game-theoretic methods are computationally intractable, while computationally tractable methods incur large utility loss.

\noindent
{\bf Our work:} We propose \emph{AttriGuard}, a practical defense against attribute inference attacks. AttriGuard is computationally tractable and incurs small utility loss. In AttriGuard, the defender's ultimate goal is to add random noise to a user's public data to minimize the attacker's inference accuracy with a small utility loss of the public data. 
Achieving this goal relies on estimating the attacker's accuracy at inferring the user's private attribute when a particular noise is added, which is challenging because 1) the defender does not know the user's true attribute value (we consider this threat model to avoid single-point failure introduced by a compromised defender), and 2) the defender does not know the attacker's classifier, since there are many possible choices for the classifier. 
To address the challenge, AttriGuard works in two phases. 

In Phase I, for each possible attribute value, the defender finds a minimum noise such that if we add the noise to the user's public data, then the attacker's classifier predicts the attribute value for the user. From the perspective of \emph{adversarial machine learning}~\cite{barreno2006can}, finding such minimum noise is known as \emph{evasion attacks} to classifiers. Specifically, in our problem, the defender adds minimum noise to evade the attacker's classifier. However, Phase I faces two challenges. The first challenge is that existing evasion attack methods~\cite{barreno2006can,Biggio:ECMLPKDD:13, Goodfellow:ICLR:14b,Papernot:arxiv:16Limitation,sharif2016accessorize,CarliniSP17} did not consider the unique characteristics of privacy protection, as they were not designed for such purpose. In particular, in defending against attribute inference attacks, different users may have different preferences on what types of noise can be added to their public data. For instance, in recommender systems, a user may prefer modifying its existing rating scores, or adding new rating scores to items the user did not rate before, or combination of them. Existing evasion attack methods did not consider such constraints. To address the challenge, we optimize an existing evasion attack, which was developed by Papernot et al.~\cite{Papernot:arxiv:16Limitation}, to incorporate such constraints. 

The second challenge is that the defender does not know the attacker's classifier. To address the challenge, the defender itself learns a classifier to perform attribute inference. Since both the attacker's classifier and the defender's classifier model the relationships between users' public data and private attributes and the two classifiers could have similar classification boundaries, the noise optimized to evade the defender's classifier is very likely to also evade the attacker's classifier. Such phenomenon is known as \emph{transferability}~\cite{Goodfellow:ICLR:14b,PracticalBlackBox17,liu2016delving} in adversarial machine learning.  
Evasion attacks are often viewed as offensive techniques. For the first time, our work shows that evasion attacks can also be used as defensive techniques. In particular, evasion attacks can play an important role at defending against attribute inference attacks.

In Phase II, the defender randomly picks an attribute value according to a probability distribution $\mathbf{q}$ over the possible attribute values and adds the corresponding noise found in Phase I to the user's public data. The probability distribution $\mathbf{q}$ roughly characterizes the probability distribution of the attacker's inference for the user. We find the probability distribution $\mathbf{q}$ via minimizing its distance to a \emph{target probability distribution $\mathbf{p}$} with a bounded utility loss of the public data. The target probability distribution is selected by the defender. For instance, the target probability distribution could be a uniform distribution over the possible attribute values, with which the defender aims to make the attacker's inference close to random guessing. Formally, we formulate finding the probability distribution $\mathbf{q}$ as solving a constrained convex optimization problem. Moreover, we develop a method based on the \emph{Karush-Kuhn-Tucker (KKT) conditions}~\cite{convexOptimization} to solve the optimization problem.

We evaluate AttriGuard and compare it with existing defenses using a real-world dataset from Gong and Liu~\cite{GongAttriInferSEC16}. In the dataset, a user's public data are the rating scores the user gave to mobile apps on Google Play, while the attribute is the city a user lives/lived in. First, our results demonstrate that our adapted evasion attack in Phase I outperforms existing ones. Second, AttriGuard is effective at defending against attribute inference attacks. For instance, by modifying at most 4 rating scores on average, the attacker's inference accuracy is reduced by 75\% for several defense-unaware attribute inference attacks and attacks that adapt to our defense. 
Third, AttriGuard adds significantly smaller noise to users' public data than existing defenses when reducing the attacker's inference accuracy by the same amount. 

In summary, our key contributions are as follows:
\begin{packeditemize}
\item We propose AttriGuard, a practical two-phase defense against attribute inference attacks. 

\item We optimize an evasion attack method to incorporate the unique characteristics of defending against attribute inference attacks in Phase I of AttriGuard. Moreover, we develop a KKT condition based solution to select the random noise in Phase II. 

\item We extensively evaluate AttriGuard and compare it with existing defenses using a real-world dataset.

\end{packeditemize}

\section{Related Work}
\label{relatedwork}

\subsection{Attribute Inference Attacks}
A number of recent studies~\cite{Otterbacher10,weinsberg2012blurme,Zheleva09,Chaabane12,kosinski2013private,Gong14,GongAttriInferSEC16,AttriInfer,Gong18tops,MichalevskyPowerSEC,Narain16Oakland,fredrikson2014privacy,fredrikson2015model,Lerman11,Zhang12} have demonstrated that users are vulnerable to \emph{attribute inference attacks}. 
In these attacks, an attacker has access to a set of measurement data about a target user, which we call \emph{public data}; 
and the attacker aims to infer \emph{private attributes} (e.g., location, political view, or sexual orientation)
of the target user. Specifically, the attacker has a machine learning classifier, which takes a user's public data as input and produces the user's attribute value. 
The classifier can be learnt on a training dataset consisting of both public data and attribute values of users who also make their attributes public. 
Next, we review several attribute inference attacks in various application domains. 

In recommender systems, a user's public data can be the list of rating scores that the user gave to certain items. 
Weinsberg et al.~\cite{weinsberg2012blurme} demonstrated that an attacker (e.g., provider of a recommender system) can use a machine learning classifier (e.g., logistic regression) to predict a user's gender
based on the user's rating scores to movies. Specifically, an attacker first collects rating scores and gender information from the users who publicly disclose both rating scores and gender; 
the attacker represents each user's rating scores as a feature vector, e.g., the $i$th entry of the feature vector is the rating score that the user gave to the $i$th movie if the user reviewed the $i$th movie, otherwise the $i$th entry is 0;
and the attacker uses the collected data as a training dataset to learn a classifier to map a user's rating scores to gender. The attacker then uses the classifier to infer gender for target users who do not disclose their gender, i.e., given a target user's rating scores, the classifier produces either male or female. 

In social media (e.g., Facebook), a user's public data could be the list of pages or musics liked or shared by the user, as well as the user's friend lists. 
Several studies~\cite{Zheleva09,Chaabane12,kosinski2013private,Gong14,GongAttriInferSEC16,AttriInfer,Gong18tops} have demonstrated that an attacker (e.g., social media provider, advertiser, or data broker) can use a machine learning classifier to infer a target user's private attributes (e.g., gender, cities lived, and political view) based on the user's public data on social media. Again, the attacker first collects a dataset from users who disclose their attributes and use them as a training dataset to learn the classifier.  The classifier is then used to infer attributes of target users who do not disclose them. 

In mobile apps, Michalevsky et al.~\cite{MichalevskyPowerSEC} showed that an attacker can use machine learning to infer a user's location based on the user's  smartphone's aggregate power consumption (i.e., ``public data" in our terminology). Narain et al.~\cite{Narain16Oakland} showed that an attacker can infer user locations using the gyroscope, accelerometer, and magnetometer
data available from the user's smartphone. In side-channel attacks~\cite{Lerman11,Zhang12}, an attacker could use power consumption and processing time 
 (i.e., public data) to infer cryptographic keys (i.e., private attribute). 

\subsection{Defenses}

\myparatight{Game-theoretic methods}
Shokri et al.~\cite{ShokriCCS12} proposed a game-theoretic method to defend against location inference attacks; the attacker performs the optimal inference attack that the attacker adapts to the defense; and the defender obfuscates the locations to protect users against the optimal inference attack. 
Calmon et al.~\cite{Calmon:2012} proposed a game-theoretic method to defend against attribute inference attacks. These methods have theoretical privacy guarantees, but they rely on optimization problems that are computationally intractable when applied to attribute inference attacks. 
Note that the method proposed by Shokri et al.~\cite{ShokriCCS12} is tractable for defending against location inference attacks, because such problem essentially has a public data vector of 1 dimension. 


\myparatight{Computationally tractable methods} 
Due to the computational challenges of the game-theoretic methods, several studies~\cite{weinsberg2012blurme,friendbasedDefense,ChenObfuscationPETS14,Salamatian:2015} proposed to develop tractable methods, with the degradation of theoretical privacy guarantees. For instance, Salamatian et al.~\cite{Salamatian:2015} proposed \emph{Quantization Probabilistic Mapping (QPM)} to approximately solve the game-theoretic optimization problem formulated by Calmon et al.~\cite{Calmon:2012}. Specifically, they cluster users' public data and use the cluster centroids to represent them. Then, they approximately solve the optimization problem using the cluster centroids. Since quantization is used, QPM has no theoretical privacy guarantee, i.e., QPM does not necessarily defend against the optimal attribute inference attacks, but QPM makes it tractable to solve the defense problem in practice. 

Other computationally tractable methods~\cite{weinsberg2012blurme,ChenObfuscationPETS14}
 leveraged heuristic correlations between the entries of the public data vector and attribute values. Specifically, they modify the $k$ entries that have large correlations with the attribute values that do not belong to the target user. $k$ is a parameter to control privacy-utility tradeoffs. 
For instance, Weinsberg et al.~\cite{weinsberg2012blurme} proposed BlurMe to defend against attribute inference attacks in the context of recommender systems. For each attribute value $i$, they order the items into a list $L_i$ according to the correlations between the items and the attribute values other than $i$. Specifically, for each attribute value $i$, they learn a logistic regression classifier via using the public data vector as a feature vector; and the negative coefficient of an item in the logistic regression classifier is treated as its correlation with the attribute values other than $i$. An item has a larger correlation means that changing the item's rating score is more likely to change the classifier's inference.
 For a target user whose attribute value is $i$, the defender selects the top-$k$ items from the list $L_i$ that were not rated by the user yet, and then adds the average rating score to those items. Chen et al.~\cite{ChenObfuscationPETS14} proposed ChiSquare, which computed correlations between items and attribute values based on chi-square statistics. 


As we elaborated in the Introduction section, these methods have one or two limitations: 1) they incur large utility loss, and 2) some of them require the defender to have direct access to users' private attribute values. 
 
%

\myparatight{Local differential privacy (LDP)} LDP~\cite{RandomizedResponse,DuchiLDP13,Erlingsson:2014,BassilySuccinctHistograms15,QinHeavyHitterCCS16,smith2017interaction,BLENDER,Wang17LDP} is a technique based on $\epsilon$-differential privacy~\cite{Dwork:2006} to protect privacy of an individual user's data record, i.e., public data in our problem. 
LDP provides a strong privacy guarantee. However, LDP aims to achieve a privacy goal that is different from the one in attribute inference attacks. 
Roughly speaking, LDP's privacy goal is to add random noise to a user's true data record such that two arbitrary true data records have close probabilities (their difference is bounded by a privacy budget)   
to generate the same noisy data record. However, in defending against attribute inference attacks, the privacy goal is to add noise to a user's public data record 
such that the user's private attributes cannot be accurately inferred by the attacker's classifier. 
As a result, as we will demonstrate in our experiments, LDP achieves a suboptimal privacy-utility tradeoff at defending against attribute inference attacks, i.e., LDP adds much larger noise than our defense to make the attacker have the same inference accuracy. 

\section{Problem Formulation}
\label{problem}

%


We have three parties: \emph{user}, \emph{attacker}, and \emph{defender}.  
The defender adds noise to a user's public data to protect its private attribute. 
Next, we discuss each party one by one.

\subsection{User} 

A user aims to publish some data while preventing inference of its private attribute from the public data. 
We denote the user's public data and private attribute as $\mathbf{x}$ (a column vector) and $s$, respectively. 
For simplicity, we assume each entry of $\mathbf{x}$ is normalized to be in the range $[0,1]$.
The  attribute $s$ has $m$ possible values, which we denote as $\{1,2,\cdots,m\}$; $s=i$ means that the user's private attribute value is $i$.
For instance, when the private attribute is political view, the attribute could have two possible values, i.e., democratic and republican. 
We note that the attribute $s$ could be a combination of multiple attributes. 
For instance, the attribute could be $s=(\text{political view, gender})$, which has four possible values, i.e., (democratic, male), (republican, male), (democratic, female), and (republican, female).

\myparatight{Policy to add noise} Different users may have different preferences over what kind of noise can be added to their public data.  For instance, in recommender systems, a user may prefer modifying its existing rating scores, 
while another user may prefer adding new rating scores.
We call a policy specifying what kind of noise can be added a \emph{noise-type-policy}. 
In particular, we consider the following three types of  noise-type-policy.
\begin{packeditemize}
\item {\bf Policy A: Modify\_Exist.} In this policy, the defender can only modify the non-zero entries of $\mathbf{x}$. In recommender systems, this policy means that the defender can only modify a user's existing rating scores; in social media, when the public data correspond to page likes, this policy means that  the defender can only remove a user's existing page likes. 
\item {\bf Policy B: Add\_New.} In this policy, the defender can only change the zero entries of $\mathbf{x}$. In recommender systems, this policy means that the defender can only add new rating scores for a user; when the public data represent page likes in social media, this policy means that  the defender can only add new page likes for a user. We call this policy \emph{Add\_New}.
\item {\bf Policy C: Modify\_Add.} This policy is a combination of Modify\_Exist and Add\_New. In particular, the defender could modify any entry of $\mathbf{x}$. 
\end{packeditemize}

\subsection{Attacker} 

The attacker has access to the noisy public data and aims to infer the user's private attribute value. 
We consider an attacker has 
 a {machine learning classifier} that takes a user's (noisy) public data as input and infers the user's private attribute value. 
Different users might treat different attributes as private. In particular, some users do not treat the attribute $s$ as private, so they publicly disclose it. 
Via collecting data from such users, the attacker can learn the machine learning classifier. 

We denote the attacker's machine learning classifier as $C_a$, and $C_a(\mathbf{x})\in$ $\{1,2,\cdots,m\}$ is the predicted attribute value for the user whose public data is $\mathbf{x}$. 
The attacker could use a standard machine learning classifier, e.g., logistic regression, random forest, and neural network. 
Moreover, an attacker can also adapt its attack based on the defense. For instance, the attacker could first try detecting the noise and then perform attribute inference attacks. We assume the attacker's classifier  is unknown to the defender, since there are many possible choices for the attacker's classifier.

\subsection{Defender} 
The defender adds noise to a user's true public data according to a noise-type-policy. The defender is a software on the user's client side.
For instance, to defend against attribute inference attacks on a social media, the defender can be an app within the social media or a browser extension.
Once a user gives privileges to the defender, the defender can modify its public data, e.g., the defender can add page likes on Facebook or rate new items in a recommender system on behalf of the user.    

The defender has access to the user's true public data $\mathbf{x}$. The defender adds a random noise vector $\mathbf{r}$ to $\mathbf{x}$, and the noise is randomly selected according to a \emph{randomized noise addition mechanism} $\mathcal{M}$. Formally, $\mathcal{M}(\mathbf{r}|\mathbf{x})$ is the probability that the defender will add noise vector $\mathbf{r}$ when the true public data is $\mathbf{x}$.
Since the defender adds random noise to the user's public data, the resulting noisy public data $\mathbf{x}+\mathbf{r}$ is a randomized vector. 
Therefore, the inference of the attacker's classifier $C_a$ is also a random variable. We denote the probability distribution of this random variable as $\mathbf{q}$, where $q_i=\text{Pr}(C_a(\mathbf{x+r})=i)$ is the probability that the classifier $C_a$ outputs $i$. 

The defender's ultimate goal is to find a mechanism $\mathcal{M}$ that minimizes the inference accuracy of the attacker's classifier with a bounded utility loss of the public data. However, the defender faces two challenges at computing such inference accuracy: 1) the defender does not know the attacker's classifier $C_a$, and 2) the defender has no access to a user's true private attribute value. Specifically, in our threat model, to avoid single-point failure introduced by a compromised defender, we consider the defender does not have direct access to the user's private attribute value. 

\myparatight{Addressing the first challenge} To address the first challenge, the defender itself learns a classifier $C$ to perform attribute inference. For instance, using the data from the users who share both public data and attribute values, the defender can learn such a classifier $C$. The defender treats the output probability distribution of the classifier $C$ as the output probability distribution $\mathbf{q}$ of the attacker's classifier. 
Moreover, we consider the defender's classifier $C$ is implemented in the popular \emph{one-vs-all} paradigm. Specifically, the classifier has $m$ decision functions denoted as $C_1$, $C_2$, $\cdots$, $C_m$, where $C_i(\mathbf{x})$ is the confidence that the user has an attribute value $i$. The classifier's inferred attribute value is  $C(\mathbf{x})=\argmax_i C_i(\mathbf{x})$. Note that, when the attribute only has two possible values (i.e., $m=2$), we have $C_2(\mathbf{x})=-C_1(\mathbf{x})$ for classifiers like logistic regression and SVM. 

\myparatight{Addressing the second challenge} To address the second challenge, we consider an alternative goal, which aims to find a mechanism $\mathcal{M}$ such that the output probability distribution $\mathbf{q}$ is the closest to a \emph{target probability distribution $\mathbf{p}$} with a utility-loss budget, where $\mathbf{p}$ is selected by the defender. For instance, without knowing anything about the attributes, the target probability distribution could be the uniform distribution over the $m$ attribute values, with which the defender aims to make the attacker's inference close to random guessing. The target probability distribution could also be estimated from the users who publicly disclose the attribute, e.g., the probability $p_i$ is the fraction of such users who have attribute value $i$. Such target probability distribution naturally represents a baseline attribute inference attack. The defender aims to reduce an attack to the baseline attack with such target probability distribution. 

The defender needs a formal metric to quantify the distance between  $\mathbf{p}$ and $\mathbf{q}$ such that the defender can find a mechanism $\mathcal{M}$ to minimize the distance.  
We measure the distance between $\mathbf{p}$ and $\mathbf{q}$ using their Kullback--Leibler (KL) divergence, i.e., $KL(\mathbf{p}||\mathbf{q})$=$\sum_i p_i\text{log}\frac{p_i}{q_i}$. We choose KL divergence because it makes our formulated optimization problem become a convex problem, which has efficient and accurate solutions. 


\myparatight{Measuring utility loss}
A user's (noisy) public data are often leveraged by a service provider to provide services. For instance,  
in a recommender system (e.g., Amazon, Google Play, Netflix), a user's public data are rating scores or likes/dislikes to items, which are used to recommend items to users that match their personalized preferences. Therefore, utility loss of the public data can essentially be measured by the service quality loss. 
Specifically, 
 in a recommender system, the decreased accuracy of the recommendations introduced by the added noise can be used as utility loss. However, using such service-dependent utility loss makes the formulated optimization problem  computationally intractable.   

 Therefore, we aim to use utility-loss metrics that make our formulated optimization problems tractable but can still well approximate the utility loss for different services. 
In particular, we can use a distance metric $d(\mathbf{x}, \mathbf{x} + \mathbf{r})$ to measure utility loss. Since $\mathbf{r}$ is a random value generated according to the mechanism $\mathcal{M}$, we will measure the utility loss using the expected distance $E(d(\mathbf{x}, \mathbf{x} + \mathbf{r}))$.  For instance, the distance metric can be $L_0$ norm of the noise, i.e., $d(\mathbf{x}, \mathbf{x} + \mathbf{r})=||\mathbf{r}||_0$. 
$L_0$ norm is the number of entries of $\mathbf{x}$ that are modified by the noise, which has semantic interpretations in a number of real-world application domains. 
For instance, in a recommender system, $L_0$ norm means the number of items whose rating scores are modified. Likewise, in social media, an entry of $\mathbf{x}$ is 1 if the user liked the corresponding page, otherwise the entry is 0. Then, $L_0$ norm means the number of page likes that are removed or added by the defender. 
The distance metric can also be $L_2$ norm of the noise, which considers the magnitude of the modified rating scores in the context of recommender systems.

\myparatight{Attribute-inference-attack defense problem} With a quantifiable defender's goal and utility loss, we can formally define the problem of defending against attribute inference attacks. Specifically, the user specifies a noise-type-policy and an utility-loss budget $\beta$. The defender specifies a target probability distribution $\mathbf{p}$, learns a classifier $C$, and finds a mechanism $\mathcal{M}^{*}$, which adds noise to the user's public data such that the user's utility loss is within the budget while the output probability distribution $\mathbf{q}$ of the classifier $C$ is closest to the target probability distribution $\mathbf{p}$. 
 Formally, we have:
\begin{definition}
Given a noise-type-policy $\mathcal{P}$, an utility-loss budget $\beta$, a target probability distribution $\mathbf{p}$, and a classifier $C$, the defender aims to find a mechanism $\mathcal{M}^{*}$ via solving the following optimization problem: 
\begin{align}
\mathcal{M}^{*}=&\argmin_{\mathcal{M}} KL(\mathbf{p}||\mathbf{q}) \nonumber \\
\label{ulb}
\text{subject to }  & E(d(\mathbf{x}, \mathbf{x} + \mathbf{r})) \leq \beta,
\end{align}
where the probability distribution $\mathbf{q}$ depends on the classifier $C$ and the mechanism $\mathcal{M}$.
\end{definition}

In this work, we use the $L_0$ norm of the noise as the metric $d(\mathbf{x}, \mathbf{x} + \mathbf{r})$ because of its semantic interpretation. 

\section{Design of AttriGuard}

\subsection{Overview}
The major challenge to solve the optimization problem in Equation~\ref{ulb} is that the number of parameters of the mechanism $\mathcal{M}$, which maps a given vector to another vector probabilistically, is exponential to the dimensionality of the public data vector.  
To address the challenge, 
we propose a \emph{two-phase framework} to solve the optimization problem. 
Our intuition is that, although the noise space is large, we can categorize them into $m$ groups depending on the defender's classifier's inference. Specifically, 
we denote by $G_i$ 
 the group of noise such that if we add any of them to the user's public data, then the defender's classifier will infer the attribute value $i$ for the user. Essentially, the probability ${q}_i$ that the defender's classifier infers attribute value $i$ for the user is the probability that $\mathcal{M}$ will produce a noise in the group $G_i$, i.e., ${q}_i=\sum_{\mathbf{r}\in G_i} \mathcal{M}(\mathbf{r}|\mathbf{x})$. AttriGuard finds one representative noise in each group and assumes $\mathcal{M}$ is a probability distribution concentrated on the representative noise.

Specifically, in Phase I, for each group $G_i$, we find a minimum noise $\mathbf{r}_i$ such that if we add $\mathbf{r}_i$ to the user's public data, then the defender's classifier predicts the attribute value $i$ for the user. We find a minimum noise in order to minimize utility loss. 
In \emph{adversarial machine learning}, this is known as \emph{evasion attack}. However, existing evasion attack methods~\cite{barreno2006can,Biggio:ECMLPKDD:13,Goodfellow:ICLR:14b,Papernot:arxiv:16Limitation,sharif2016accessorize,CarliniSP17} are insufficient to find the noise $\mathbf{r}_i$ in our problem, because they do not consider the noise-type-policy. We optimize an existing evasion attack method developed by Papernot et al.~\cite{Papernot:arxiv:16Limitation} to incorporate noise-type-policy. 
The noise $\mathbf{r}_i$ optimized to evade the defender's classifier is also very likely to make the attacker's classifier predict the attribute value $i$ for the user, which is known as \emph{transferability}~\cite{Goodfellow:ICLR:14b,PracticalBlackBox17,liu2016delving} in adversarial machine learning. 

In Phase II, we simplify the mechanism $\mathcal{M^*}$ to be a probability distribution over the $m$ representative noise $\{\mathbf{r}_1, \mathbf{r}_2, \cdots, \mathbf{r}_m\}$. In other words, the defender randomly samples a noise $\mathbf{r}_i$ according to the probability distribution $\mathcal{M^*}$ and adds the noise to the user's public data. Under such simplification, $\mathcal{M^*}$ only has at most $m$ non-zero parameters, the output probability distribution $\mathbf{q}$ of the defender's classifier essentially becomes $\mathcal{M^*}$, and we can transform the optimization problem in Equation~\ref{ulb} to be a convex problem. 
 Moreover, we design a method based on the \emph{Karush-Kuhn-Tucker (KKT) conditions}~\cite{convexOptimization} to solve the convex optimization problem. 

\subsection{Phase I: Finding $\mathbf{r}_i$}

The user's public data is $\mathbf{x}$. Suppose we aim to add a minimum noise $\mathbf{r}_i$ to $\mathbf{x}$, according to the noise-type-policy $\mathcal{P}$, 
such that the classifier $C$ infers the attribute value $i$ for the user. Formally, we model finding such  $\mathbf{r}_i$ as solving the following optimization problem:
\begin{align}
& \mathbf{r}_i = \argmin_{\mathbf{r}} ||\mathbf{r}||_0 \nonumber \\
\label{findr}
\text{subject to } & C(\mathbf{x+r})=i.
\end{align}

\begin{algorithm}[t]
\caption{Policy-Aware Noise Finding Algorithm}
\begin{algorithmic}[1]
\REQUIRE Public data $\textbf{x}$, classifier $\textbf{C}$, noise-type-policy $\mathcal{P}$, target attribute value $i$, and step size $\tau$. \\
\ENSURE  Noise $\textbf{r}_i$. \\
    Initialize $t=0,\overline{\textbf{x}}=\textbf{x}$. \;
	
	\WHILE { $\textbf{C}(\overline{\textbf{x}}) \neq i$ and $t \leq \text{maxiter}$} \;
	\STATE //Find the entry to be modified. \;
	\IF{$\mathcal{P}==Add\_New$} \;
	
	\STATE $e_{inc}=\argmax_{j} \{\frac{\partial \textbf{C}_{i}(\overline{\mathbf{x}})}{\partial \mathbf{x}_j}| \mathbf{x}_j=0\} $ \;
	\ENDIF\;
	
	\IF{$\mathcal{P}==Modify\_Exist$} \;
	\STATE $e_{inc}=\argmax_{j} \{(1-\overline{\textbf{x}}_j) \frac{\partial \textbf{C}_{i}(\overline{\mathbf{x}})}{\partial \mathbf{x}_j}| \mathbf{x}_j \neq 0\}$ \;
	
	\STATE $e_{dec}=\argmax_{j} \{-\overline{\textbf{x}}_j \frac{\partial \textbf{C}_{i}(\overline{\mathbf{x}})}{\partial \mathbf{x}_j}| \mathbf{x}_j \neq 0\}$ \;
	\ENDIF \\
	
	\IF{$\mathcal{P}==Modify\_Add$} \;
	\STATE $e_{inc}=\argmax_{j} \{(1-\overline{\textbf{x}}_j) \frac{\partial \textbf{C}_{i}(\overline{\mathbf{x}})}{\partial \mathbf{x}_j}\}$ \;
	
	\STATE $e_{dec}=\argmax_{j} \{-\overline{\textbf{x}}_j \frac{\partial \textbf{C}_{i}(\overline{\mathbf{x}})}{\partial \mathbf{x}_j}\}$ \;
	\ENDIF \\

	\STATE //Modify the entry $\overline{\textbf{x}}_{e_{inc}}$ or $\overline{\textbf{x}}_{e_{dec}}$ depending on which one is more beneficial. \;
	\STATE $v_{inc}=(\textbf{1}-\overline{\textbf{x}}_{e_{inc}}) \frac{\partial \textbf{C}_{i}(\overline{\mathbf{x}})}{\partial \mathbf{x}_{e_{inc}}}$ \;
	\STATE $v_{dec}=-\overline{\textbf{x}}_{e_{dec}} \frac{\partial \textbf{C}_{i}(\overline{\mathbf{x}})}{\partial \mathbf{x}_{e_{dec}}}$ \;
	\IF {$\mathcal{P}==Add\_New$ or $v_{inc} \geq  v_{dec}$} \;
	
	\STATE $\overline{\textbf{x}}_{e_{inc}} = clip(\overline{\textbf{x}}_{e_{inc}}+ \tau)$ \;
	\label{clip1}
	\ELSE 
	
	\STATE $\overline{\textbf{x}}_{e_{dec}} = clip(\overline{\textbf{x}}_{e_{dec}} - \tau)$ \;
	\label{clip2}
	\ENDIF \;
	\STATE $t=t+1$ \;
	\ENDWHILE \;
	\RETURN $\overline{\textbf{x}}-\textbf{x}$. \;

\end{algorithmic}
\label{algorithm2}
\end{algorithm}

Our formulation of finding $\mathbf{r}_i$ is closely related to \emph{adversarial machine learning}. 
In particular, finding $\mathbf{r}_i$ can be viewed as an \emph{evasion attack}~\cite{barreno2006can,Biggio:ECMLPKDD:13,Goodfellow:ICLR:14b,Papernot:arxiv:16Limitation,sharif2016accessorize,CarliniSP17} to the classifier $C$. 
However, existing evasion attack algorithms (e.g.,~\cite{Goodfellow:ICLR:14b,Papernot:arxiv:16Limitation,CarliniSP17}) are insufficient to solve $\mathbf{r}_i$ in our problem. The key reason is that they do not consider the noise-type-policy, which specifies the types of noise that can be added. 
We note that evasion attacks to machine learning are generally treated as offensive techniques, but our work demonstrates that evasion attacks can also be used as defensive techniques, e.g., defending against attribute inference attacks. 

Papernot et al.~\cite{Papernot:arxiv:16Limitation} proposed a \emph{Jacobian-based Saliency Map  Attack} (JSMA) to deep neural networks.  They demonstrated that JSMA can find small noise (measured by $L_0$ norm) to evade a deep neural network.  Their algorithm iteratively adds noise to an example ($\mathbf{x}$ in our case) until the classifier $C$ predicts $i$ as its label or the maximum number of iterations is reached. 
In each iteration, the algorithm picks  one or two entries of $\mathbf{x}$ based on saliency map, and then increase or decrease the entries by a {constant} value. 

We also design our algorithm based on saliency map. However, our algorithm is different from JSMA in two aspects.  First, our algorithm incorporates the noise-type-policy, while theirs does not. The major reason is that their algorithm is not developed for preserving privacy, so they do not have noise-type-policy as an input. 
Second, in their algorithm, all the modified entries of  $\mathbf{x}$ are either increased or decreased. In our algorithm, some entries can be increased while other entries can be decreased. 
As we will demonstrate in our experiments, our algorithm can find smaller noise than JSMA.

Algorithm~\ref{algorithm2} shows our algorithm to find $\mathbf{r}_i$. We call our algorithm \emph{\underline{P}olicy-\underline{A}ware \underline{N}oise Fin\underline{d}ing \underline{A}lgorithm (PANDA)}. Roughly speaking, in each iteration, based on the noise-type-policy and saliency map, we find the entry of $\mathbf{x}$, by increasing or decreasing which the noisy public data could most likely move towards the class $i$. Then, we modify the entry by $\tau$, which is a parameter in our algorithm. We will discuss setting $\tau$ in our experiments. The operation $clip(y)$ at lines~\ref{clip1} and~\ref{clip2} normalizes the value $y$ to be in [0,1], i.e., $clip(y)=1$ if $y>1$,  $clip(y)=0$ if $y<0$, and $clip(y)=y$ otherwise. 
We note that, for the noise-type-policy Modify\_Add, our algorithm can always find a solution $\mathbf{r}_i$, because this policy allows us to explore each possible public data vector. However, for the policies Modify\_Exist and Add\_New, there might exist no solution $\mathbf{r}_i$ for the optimization problem in Equation~\ref{findr}. In such cases, we will automatically extend to the Modify\_Add policy.

\subsection{Phase II: Finding $\mathcal{M}^*$}
In AttriGuard, after the defender solves $\{\mathbf{r}_1, \mathbf{r}_2, \cdots, \mathbf{r}_m\}$, the defender randomly samples one of them with a certain probability and adds it to the user's public data $\mathbf{x}$.
Therefore, in our framework, the randomized noise addition mechanism $\mathcal{M}$ is a probability distribution over $\{\mathbf{r}_1, \mathbf{r}_2, \cdots, \mathbf{r}_m\}$, where $\mathcal{M}_i$ is the probability that the defender adds $\mathbf{r}_i$ to $\mathbf{x}$. 
Since $q_i=\text{Pr}(C(\mathbf{x+r})=i)$ and $C(\mathbf{x+r}_i)=i$, we have $q_i=\mathcal{M}_i$, where $i\in \{1,2,\cdots,m\}$. 
Therefore, we can transform the optimization problem in Equation~\ref{ulb} to the following optimization problem: 
\begin{align}
\mathcal{M}^{*}=&\argmin_{\mathcal{M}} KL(\mathbf{p}||\mathcal{M}) \nonumber \\
\label{ulb-1}
\text{subject to }  & \sum_{i=1}^m \mathcal{M}_i ||\mathbf{r}_i||_0 \leq \beta \nonumber \\
& \mathcal{M}_i > 0, \forall i\in \{1,2,\cdots,m\} \nonumber \\
& \sum_{i=1}^m \mathcal{M}_i = 1,
\end{align}
where we use the $L_0$ norm of the noise as the utility-loss metric $d(\mathbf{x}, \mathbf{x} + \mathbf{r})$ in Equation~\ref{ulb}. 

Next, we discuss how to solve the above optimization problem. 
 We can show that the above optimization problem is convex because its objective function and constraints are convex, which implies that $\mathcal{M}^*$ is a global minimum.  Therefore, according to the standard \emph{Karush-Kuhn-Tucker (KKT) conditions}~\cite{convexOptimization}, we have the following equations:
\begin{align}
\label{kkt11}
&\triangledown_{\mathcal{M}}(KL(\mathbf{p}||\mathcal{M}^{*}) + \mu_0 (\sum_{i=1}^m \mathcal{M}_i^{*} ||\mathbf{r}_i||_0 - \beta) - \sum_{i=1}^m \mu_{i}\mathcal{M}_i^{*} \nonumber \\
& + \lambda (\sum_{i=1}^m \mathcal{M}_i^{*} -1))=0  \\
\label{kkt12}
&\mu_{i}\mathcal{M}_i^{*} = 0, \forall i\in \{1,2,\cdots,m\}  \\
\label{kkt13}
&\mu_0 (\sum_{i=1}^m \mathcal{M}_i^{*} ||\mathbf{r}_i||_0 - \beta) =0,
\end{align}
where $\triangledown$ indicates gradient, while $\mu_i$ and $\lambda$ are KKT multipliers.  
Then, we can obtain the following equations:
\begin{align}
\label{prb-1-3}
&\mu_i = 0, \forall i\in \{1,2,\cdots,m\} \\
\label{ex1}
&\mathcal{M}_{i}^{*}=\frac{{p}_{i}}{\mu_0 ||\mathbf{r}_i||_0 + \lambda } \\
\label{ex2}
&\sum_{i=1}^m \mathcal{M}_i^{*} ||\mathbf{r}_i||_0 - \beta = 0 \\
\label{ex3}
&\mu_0=\frac{1-\lambda}{\beta}.
\end{align}

We briefly explain how we obtain Equations~\ref{prb-1-3}-\ref{ex3} from the KKT conditions. First, according to Equation~\ref{kkt12} and $\mathcal{M}_i^*>0$, we have Equation~\ref{prb-1-3}. Then, according to Equation~\ref{kkt11} and Equation~\ref{prb-1-3}, we have Equation~\ref{ex1}. Moreover, we have Equation~\ref{ex2} from Equation~\ref{kkt13} since $\mu_0\neq 0$. Finally, since $\sum_{i=1}^m \mathcal{M}_i^* = 1$, we further have Equation~\ref{ex3} from Equation~\ref{ex1} and Equation~\ref{ex2}.

Via substituting $\mathcal{M}_{i}^{*}$ in Equation~\ref{ex2} with Equation~\ref{ex1} and Equation~\ref{ex3}, we obtain a nonlinear equation with a single variable $\lambda$. We can use the Newton's method to solve $\lambda$, and then we can obtain $\mu_0$ in Equation~\ref{ex3} and $\mathcal{M}^{*}$ from Equation~\ref{ex1}.

\myparatight{Interpreting our mechanism $\mathcal{M}^*$} If we do not have the utility-loss constraint $\sum_{i=1}^m \mathcal{M}_i ||\mathbf{r}_i||_0 \leq \beta$ in the optimization problem in Equation~\ref{ulb-1}, then the mechanism $\mathcal{M}^*=\mathbf{p}$ reaches the minimum KL divergence $KL(\mathbf{p}||\mathcal{M})$, where $\mathbf{p}$ is the target probability distribution selected by the defender. In other words, if we do not consider utility loss, the defender samples the noise $\mathbf{r}_i$ with the target probability $p_i$ and adds it to the user's public data. However, when we consider the utility-loss budget, the relationship between the mechanism $\mathcal{M}^*$ and the target probability distribution $\mathbf{p}$ is represented in Equation~\ref{ex1}. In other words, the defender samples the noise $\mathbf{r}_i$ with a probability that is the target probability $p_i$ normalized by the magnitude of the noise $\mathbf{r}_i$.

\section{Evaluations}
\label{exp}

\subsection{Experimental Setup}

\subsubsection{Dataset} 
\begin{figure}[!t]
\centering
\includegraphics[width=0.38 \textwidth]{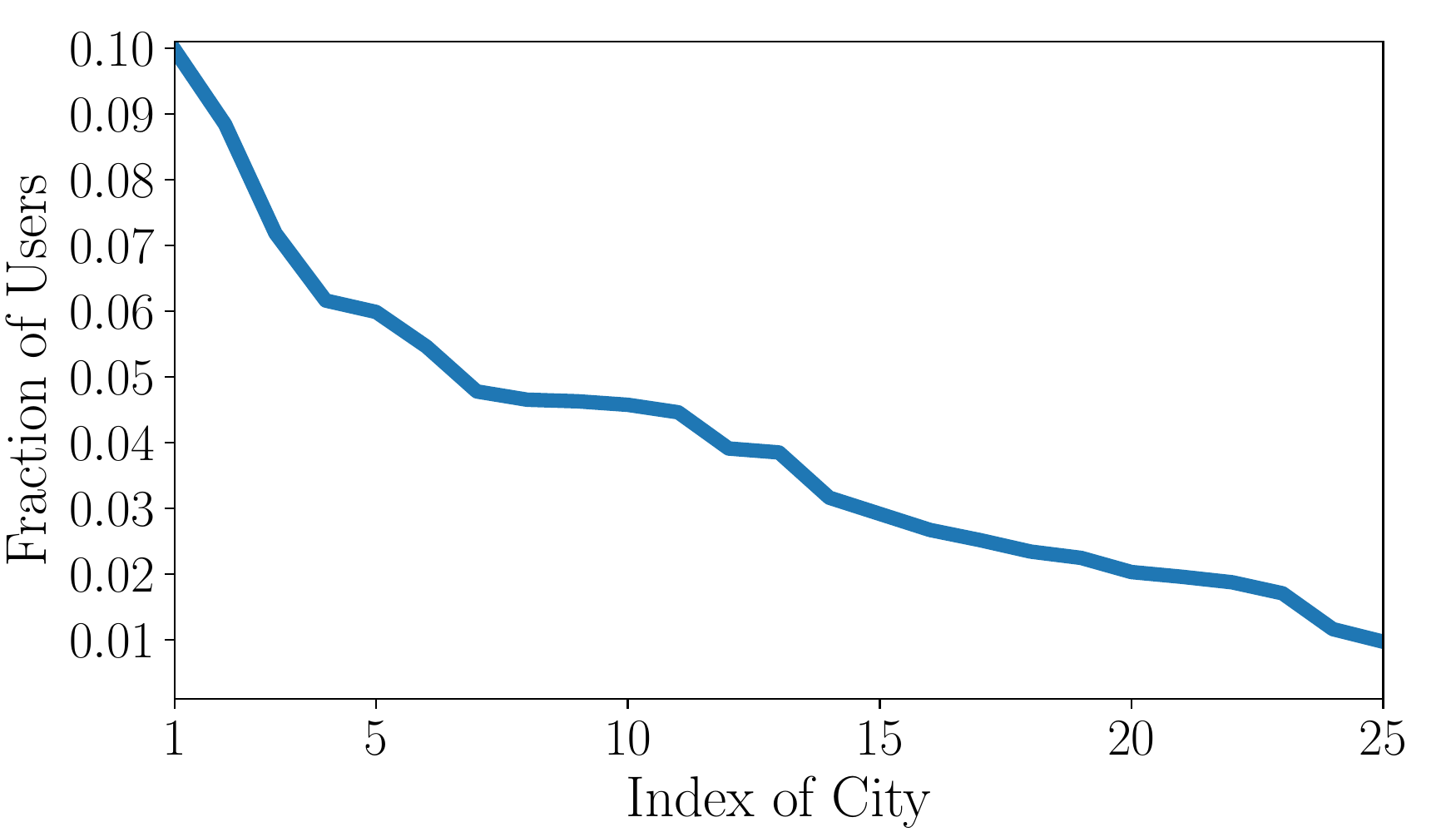}
\caption{Fraction of users who live/lived in a city.}
\vspace{-4mm}
\label{citypopu}
\end{figure}

We obtained a review dataset from Gong and Liu~\cite{GongAttriInferSEC16}. The public data of a user are the Google Play apps the user rated. We selected 10,000 popular apps and kept the users who reviewed at least 10 apps. 
In total, we have 16,238 users, and each user rated 23.2 apps on average. We represent a user's public data as a 10,000-dimension vector $\mathbf{x}$, where each entry corresponds to an app. If the user rated an app, the corresponding entry is the rating score (i.e., 1, 2, 3, 4, or 5),  
otherwise the corresponding entry has a value of 0. The attribute is the city a user lives/lived in, which were collected from users' Google+ profiles and obtained from Gong et al.~\cite{Gong12-imc}.  In total, we consider 25 popular cities. Figure~\ref{citypopu} shows the fraction of users that live/lived in a particular city. Note that we normalize each entry of a user's public data vector (i.e., review data vector) to be in [0,1], i.e., each entry is 0, 0.2, 0.4, 0.6, 0.8, or 1.0.
 
\myparatight{Training and testing} We sample 90\% of the users in the dataset uniformly at random and assume that they publicly disclose their cities lived, e.g., on Google+. The app review data and lived cities of these users are called \emph{training dataset}.  The remaining users do not disclose their cities lived, and we call them \emph{testing dataset}.


\subsubsection{Attribute Inference Attacks} 
An attribute inference attack aims to infer the cities lived for the testing users. Specifically, an attacker learns a multi-class classifier, which takes a review data vector as an input and infers the city lived, using the training dataset. We evaluate an attack using the \emph{inference accuracy} of the classifier used by the attack. Formally, 
 the {inference accuracy} of a classifier is the fraction of testing users that the inferred city lived is correct.
Since the defender does not know the attacker's classifier, we evaluate the effectiveness of AttriGuard against various attribute inference attacks as follows (we use a suffix ``-A" to indicate the classifiers are used by the attacker): 

\myparatight{Baseline attack (BA-A)} In this baseline attack, the attacker computes the most popular city among the users in the training dataset. The attacker predicts the most popular city for every user in the testing dataset. The inference accuracy of this baseline attack will not be changed by defenses that add noise to the testing users.

\myparatight{Logistic regression (LR-A)} In this attack, the attacker uses a multi-class logistic regression classifier to perform attribute inference attacks. The LR classifier was also used by previous attribute inference attacks~\cite{kosinski2013private,weinsberg2012blurme,GongAttriInferSEC16,AttriInfer}.


\myparatight{Random forest (RF-A)} In this attack, the attacker uses a random forest classifier to perform attacks.  

\myparatight{Neural network (NN-A)}  We consider the attacker uses a three-layer (i.e., input layer, hidden layer, and output layer) fully connected neural network to perform attacks. 
The hidden layer has 30,000 neurons. The output layer is a softmax layer.
We adopt the \emph{rectified linear units} as the activation function for neurons as it was demonstrated to outperform other activation functions~\cite{ReLU}. 
Note that the three-layer NN-A classifier might not be the best neural network classifier for inferring the city lived. However, exploring the best NN-A is not the focus of our work.


\myparatight{Robust classifiers: adversarial training (AT-A), defensive distillation (DD-A), and region-based classification (RC-A)} Since our defense AttriGuard leverages evasion attacks to find the noise, an attacker could leverage classifiers that are more robust to evasion attacks, based on the knowledge of our defense. We consider robust classifiers based on adversarial training~\cite{Goodfellow:ICLR:14b}, defensive distillation~\cite{Papernot16Distillation}, and region-based classification~\cite{region}. In adversarial training, an attacker generates noise for each user in the training dataset using AttriGuard and learns the neural network classifier NN-A using the noisy training dataset. In defensive distillation, an attacker refines its neural network classifier NN-A using soft labels. In region-based classification, for each testing user with a certain review data vector, an attacker randomly samples $n$ data points from a hypercube centered at the review data vector; applies the NN-A classifier to predict the attribute for each sampled data point; and the attacker takes a majority vote among the sampled data points to infer the user's attribute. We set $n=100$.  

\myparatight{Detecting noise via low-rank approximation (LRA-A)} An attacker could detect noise, remove the noise, and then perform attribute inference attacks. Whether the noise added by AttriGuard can be detected by an attacker and how to detect it effectively are not the focuses of this work, though we believe they are interesting future works. In this work, we try one way of detecting noise. 
An attacker essentially obtains a matrix of (noisy) public data for users, where each row corresponds to a user. Each entry of the matrix is a rating score or 0 if the corresponding user did not rate the item. It was well known that, in recommender systems, a normal rating-score matrix can be explained by a small number of latent factors. Therefore, 
an attacker could perform a \emph{low-rank approximation (LRA)} of the matrix. After low-rank approximation, each row could be viewed as the de-noised rating scores of a user. Then, the attacker uses these de-noised rating scores to learn a classifier NN-A and uses it to perform attribute inference.      
We implemented LRA using non-negative matrix factorization with a rank 500.


\begin{table}[!t]\renewcommand{\arraystretch}{1}
\centering
\caption{Inference accuracy of different attribute inference attacks when no defense is used.}
\begin{tabular}{|c|c|} \hline 
Attack & Inference Accuracy \\ \hline
BA-A & 0.10 \\ \hline
LR-A & 0.43 \\ \hline
RF-A & 0.44 \\ \hline
NN-A & 0.39 \\ \hline
AT-A & 0.39 \\ \hline
DD-A & 0.40 \\ \hline
RC-A & 0.38 \\ \hline
LRA-A & 0.27\\ \hline
\end{tabular} 
\label{inferAcc} 
\end{table}

\begin{figure}[!t]
\centering
{\includegraphics[width=0.4 \textwidth]{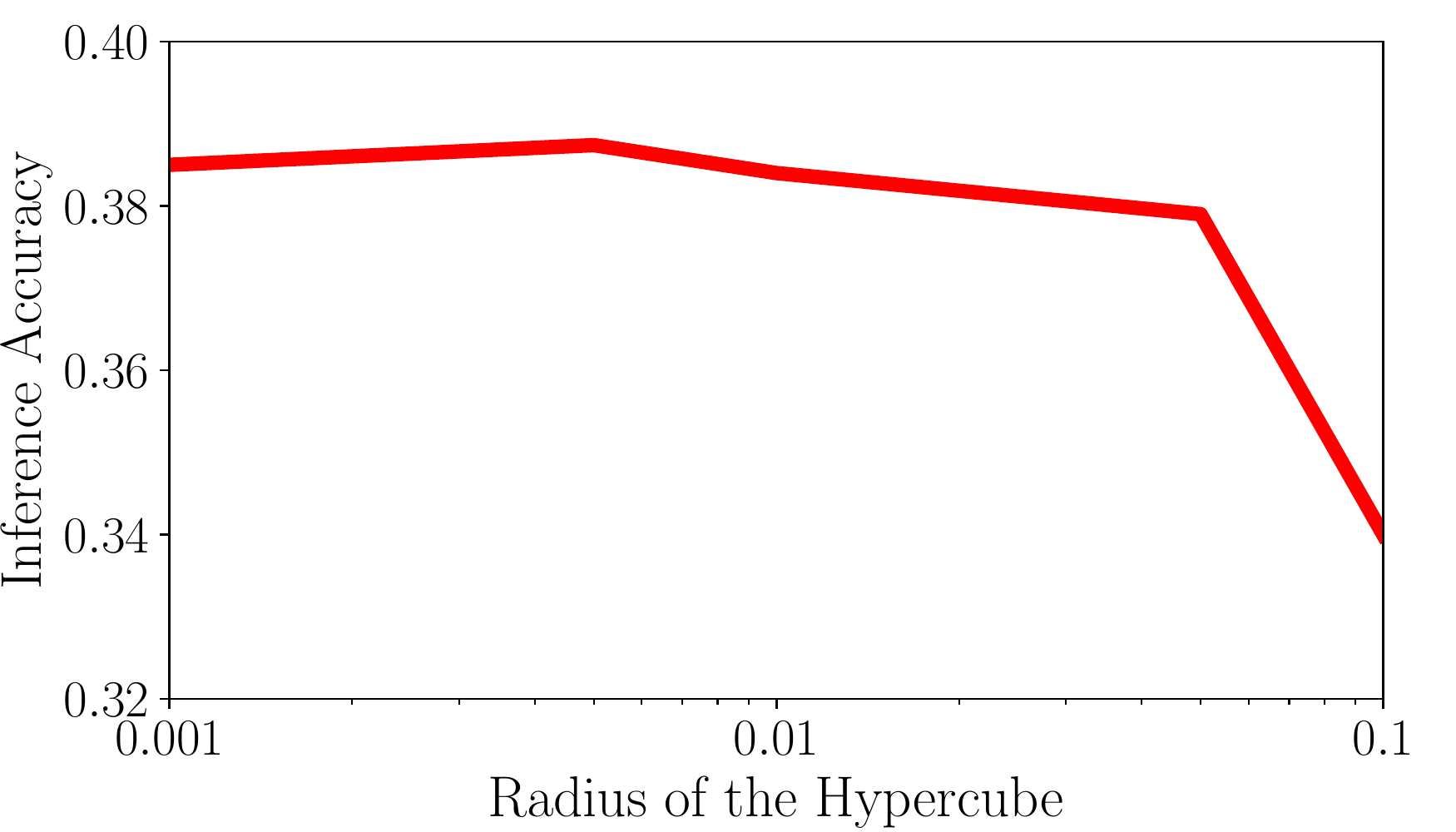}}
\caption{Inference accuracy vs. radius of the hypercube for RC-A.}
\vspace{-2mm}
\label{rc}
\end{figure}

The attacks BA-A, LR-A, RF-A, and NN-A are unaware of the defense, while AT-A, DD-A, RC-A, and LRA-A are attacks that adapt to defense. 
Table~\ref{inferAcc} shows the inference accuracy of each attack for the testing users when no defense is used. 
We note that RC-A's inference accuracy depends on the radius of the hypercube. Figure~\ref{rc} shows the inference accuracy as a function of the radius for RC-A. 
After 0.05, the inference accuracy drops sharply. Therefore, we set the radius to be 0.05 in our experiments (we use a relatively large radius to be more robust to noise added to the review data vectors). 

Without otherwise mentioned, we assume the attacker uses NN-A because it is harder for the defender to guess the neural network setting. 
 Gong and Liu~\cite{GongAttriInferSEC16} proposed an attribute inference attack. However, their attack requires both social friends and behavior data. Since our work focuses on attribute inference attacks that only use behavior data (i.e., app review data in our experiments), we do not compare with their attack.

 \begin{figure*}[!t]
\centering
\subfloat[LR-D]{\includegraphics[width=0.45 \textwidth]{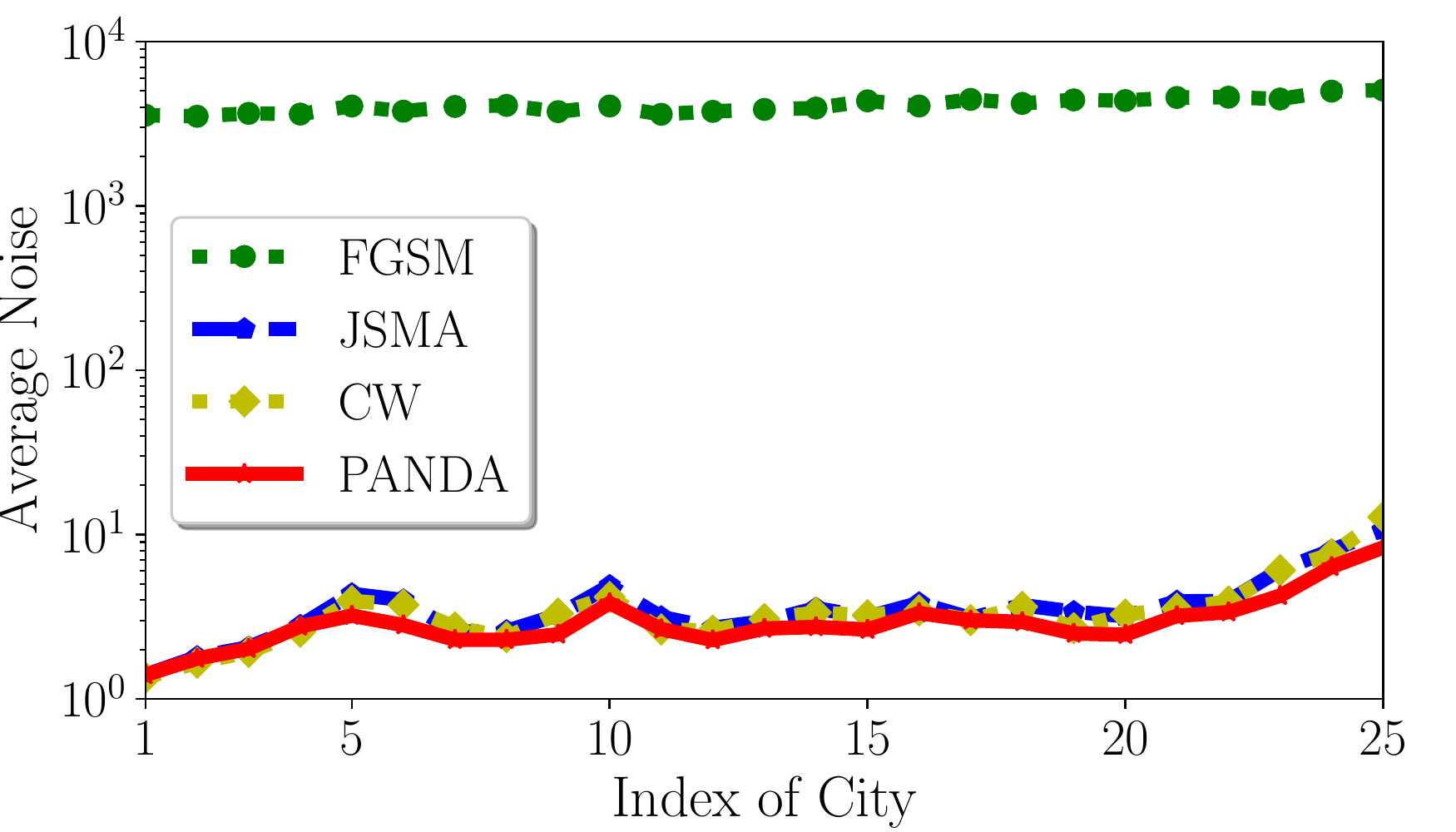}}
\subfloat[NN-D]{\includegraphics[width=0.45 \textwidth]{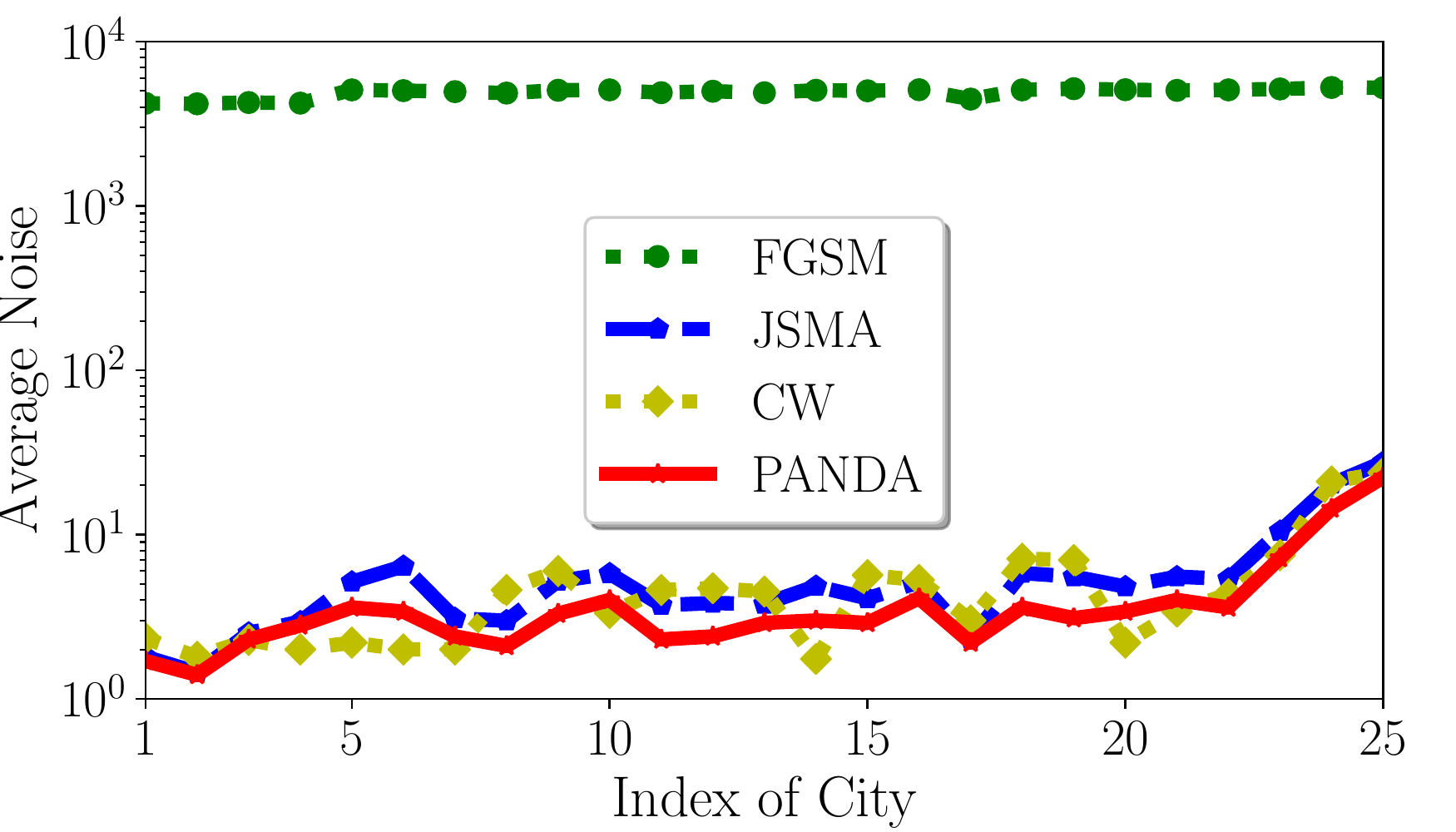}}
\caption{Average noise for each city. The defender's classifier is (a) LR-D and (b) NN-D, respectively.}
\label{noise}
\end{figure*}

\subsubsection{Parameter Setting in AttriGuard} 
The defender aims to leverage our AttriGuard to protect the cities lived for the testing users. 

\myparatight{Target probability distribution $\mathbf{p}$} We consider two possible target probability distributions. 
\begin{packeditemize}
\item {\bf Uniform probability distribution $\mathbf{p}_u$.} Without any information about the cities lived, the target probability distribution (denoted as $\mathbf{p}_u$) could be the uniform probability distribution over the 25 cities, with which the defender aims to minimize the difference between an attacker's inference and random guessing subject to a utility-loss budget. 
\item {\bf Training-dataset-based $\mathbf{p}_t$.} When the defender has access to the data of some users (e.g., users in the training dataset) who publicly disclose their cities, the defender can estimate the target probability distribution (denoted as $\mathbf{p}_t$) from such data. Specifically, the target probability for city $i$ is the fraction of training users who have city $i$. With such target probability distribution, the defender aims to minimize the difference between an attacker's inference and the baseline attack BA-A. 
\end{packeditemize}

Without otherwise mentioned, we assume the defender uses the second target probability distribution $\mathbf{p}_t$ since it considers certain knowledge about the attributes.

\myparatight{Defender's classifier $C$ (LR-D and NN-D)}  We consider two choices for the defender's classifier, i.e., multi-class logistic regression (LR-D) and neural network (NN-D). 
To distinguish between the classifiers used by the attacker and those used by the defender, we use a suffix ``-A" for each attacker's classifier while we use a suffix ``-D" for a defender's classifier.  
We note that the defender could choose any differentiable classifier. We require differentiable classifiers because our evasion attack algorithm PANDA in Phase I is applicable to differentiable classifiers. For the NN-D classifier, we also consider a three-layer fully connected neural network. However, unlike NN-A that is used by the attacker, we assume the hidden layer of the NN-D classifier has 50,000 neurons. Without otherwise mentioned, we assume the defender uses the LR-D classifier and learns it using the training dataset. We adopt LR-D as the default classifier because it is much more efficient to generate noise in Phase I. We will study the effectiveness of our defense when the attacker and the defender use different dataset to learn their classifiers.  

\myparatight{Other parameters} We set $\tau$ in our algorithm PANDA to be 1.0 when finding the minimum noise. 
Without otherwise mentioned, we set the noise-type-policy to be Modify\_Add.

 \begin{table}[!t]\renewcommand{\arraystretch}{1}
\centering
\caption{Average success rates and running times.}
\begin{tabular}{|c|c|c|c|c|} \hline 
 \multirow{2}{*}{Method} & \multicolumn{2}{c|}{Success Rate} & \multicolumn{2}{c|}{Running Time (s)} \\ \cline{2-5} 
&LR-D & NN-D & LR-D & NN-D \\ \hline
{ {FGSM}} & {  100\%} & 100\% &  {7.6} & 84 \\ \hline
{ {JSMA}} & {  100\%}& {  100\%} & {9.0}& {295}  \\ \hline
{ {CW}} & { 75\%} & { 71\%} & {7,406} & {1,067,610} \\ \hline
{ {PANDA}} & { 100\%} & { 100\%} & {8.7} & {272}  \\ \hline
\end{tabular} 
\label{comp_method} 
\end{table}

 \subsection{Results}

\myparatight{Comparing PANDA with existing evasion attack methods} 
We compare PANDA with the following evasion attack methods at finding the noise $\mathbf{r}_i$ in Phase I: \emph{Fast Gradient Sign Method} (FGSM)~\cite{Goodfellow:ICLR:14b}, \emph{Jacobian-based Saliency Map Attack} (JSMA)~\cite{Papernot:arxiv:16Limitation}, and \emph{Carlini and Wagner Attack} (CW)~\cite{CarliniSP17}. We leveraged the open-source implementation of CW published by its authors. 
The CW attack has three variants that are optimized to find small noise measured by $L_0$, $L_2$, and $L_\infty$ norms, respectively. We use the one that optimizes $L_0$ norm.
We focus on the noise-type-policy Modify\_Add, because FGSM, JSMA, and CW are not applicable to other policies. 
Note that after a method produces a noise $\mathbf{r}_i$, we will round each entry to be 0, 0.2, 0.4, 0.6, 0.8, or 1.0 since our noisy public data are discrete rating scores, and the rounded $\mathbf{r}_i$ is treated as the final noise.

Figure~\ref{noise} shows their noise (measured by $L_0$ norm) averaged over test users for each city. Moreover, Table~\ref{comp_method} shows the \emph{success rate} and \emph{running time} averaged over test users for each compared method. For each method, a test user's success rate is the fraction of cities for which the method can successfully find a $\mathbf{r}_i$ to make the classifier infer the $i$th city for the test user, and a test user's running time is the time required for the method to find $\mathbf{r}_i$ for all cities. We set the step size parameter $\epsilon$ in FGSM to be 1 as we aim to achieve a high success rate. Note that the value of $\epsilon$ does not impact the $L_0$ norm of the noise generated by FGSM.

\begin{figure}[!t]
\centering
{\includegraphics[width=0.4 \textwidth]{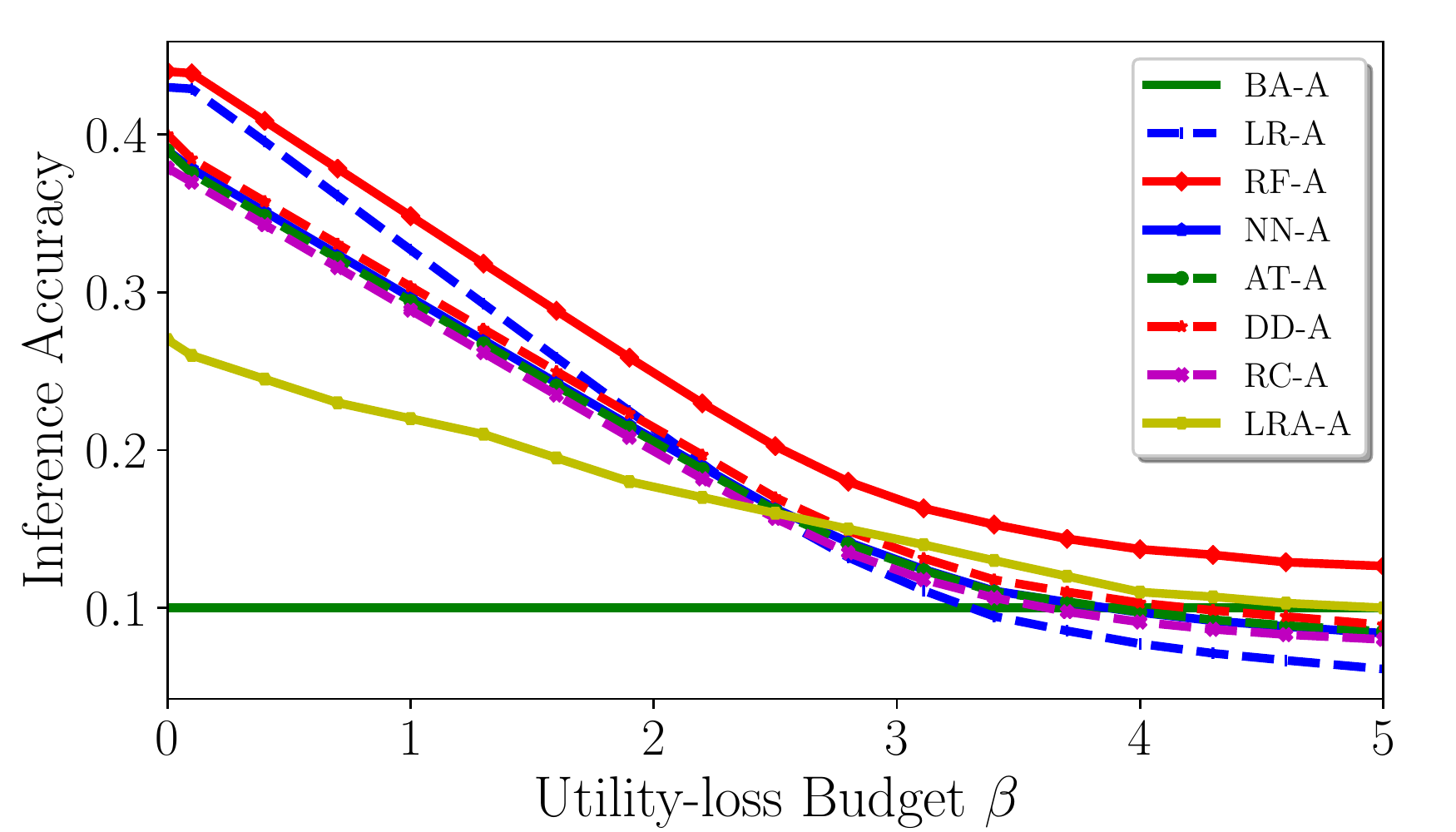}}
\caption{Attacker's inference accuracy vs. utility-loss budget.}
\label{up}
\end{figure}

First, FGSM adds orders of magnitude larger noise than other methods. This is because FGSM aims to minimize noise with respect to $L_\infty$ norm instead of $L_0$ norm.
Second, PANDA adds smaller noise and is slightly faster than JSMA for both LR-D and NN-D classifiers. This is because PANDA allows more flexible noise, i.e., some entries can be increased while other entries can be decreased in PANDA, while all modified entries can either be increased or decreased in JSMA. PANDA is faster than JSMA because it adds smaller noise and thus it runs for less iterations. Third, PANDA adds no larger noise than CW for the LR-D classifier; and PANDA adds smaller noise for some cities, but larger noise for other cities for the NN-D classifier. However, CW only has success rates less than 80\%, because of rounding the noise to be consistent with rating scores.  
 Moreover, PANDA is around 800 times and 4,000 times faster than CW for the LR-D and NN-D classifiers, respectively.  
 Considering the tradeoffs between the added noise, success rate, and running time, we recommend to use PANDA for finding noise in Phase I of AttriGuard. 

We note that JSMA and CW have similar noise for the LR-D classifier, and CW even has larger noise than JSMA for certain cities for the NN-D classifier.  Carlini and Wagner~\cite{CarliniSP17} found that CW outperforms JSMA. We suspect the reason is that our results are on review data, while their results are about image data.

\myparatight{Effectiveness of AttriGuard} Figure~\ref{up} shows the inference accuracy of various attribute inference attacks as the utility-loss budget increases, where the defender's classifier is LR-D. AttriGuard is effective at defending against attribute inference attacks. For instance, when modifying 3-4 rating scores on average, several attacks become less effective than the baseline attack. The inference accuracy of LR-A decreases the fastest as the utility-loss budget increases. This is because the defender uses LR-D, and the noise optimized based on LR-D is more likely to transfer to LR-A. The adversarial training attack AT-A has almost the same inference accuracy as NN-A. The reason is that adversarial training is not robust to iterative evasion attack~\cite{AT17} and PANDA is an iterative evasion attack. Defensive distillation attack DD-A has slightly higher inference accuracies than NN-A, because defensive distillation is more robust to the saliency map based evasion attacks~\cite{Papernot16Distillation}. LRA-A is more robust to the noise added by AttriGuard, i.e., the inference accuracy of LRA-A decreases the slowest as the utility-loss budget increases and LRA-A has higher inference accuracies than other attacks except RF-A when the utility-loss budget is larger than 3. However, AttriGuard is still effective against LRA-A since LRA-A still has low  inference accuracies and approaches to the baseline attack as the utility-loss budget increases. 

\begin{figure}[!t]
\centering
{\includegraphics[width=0.4 \textwidth]{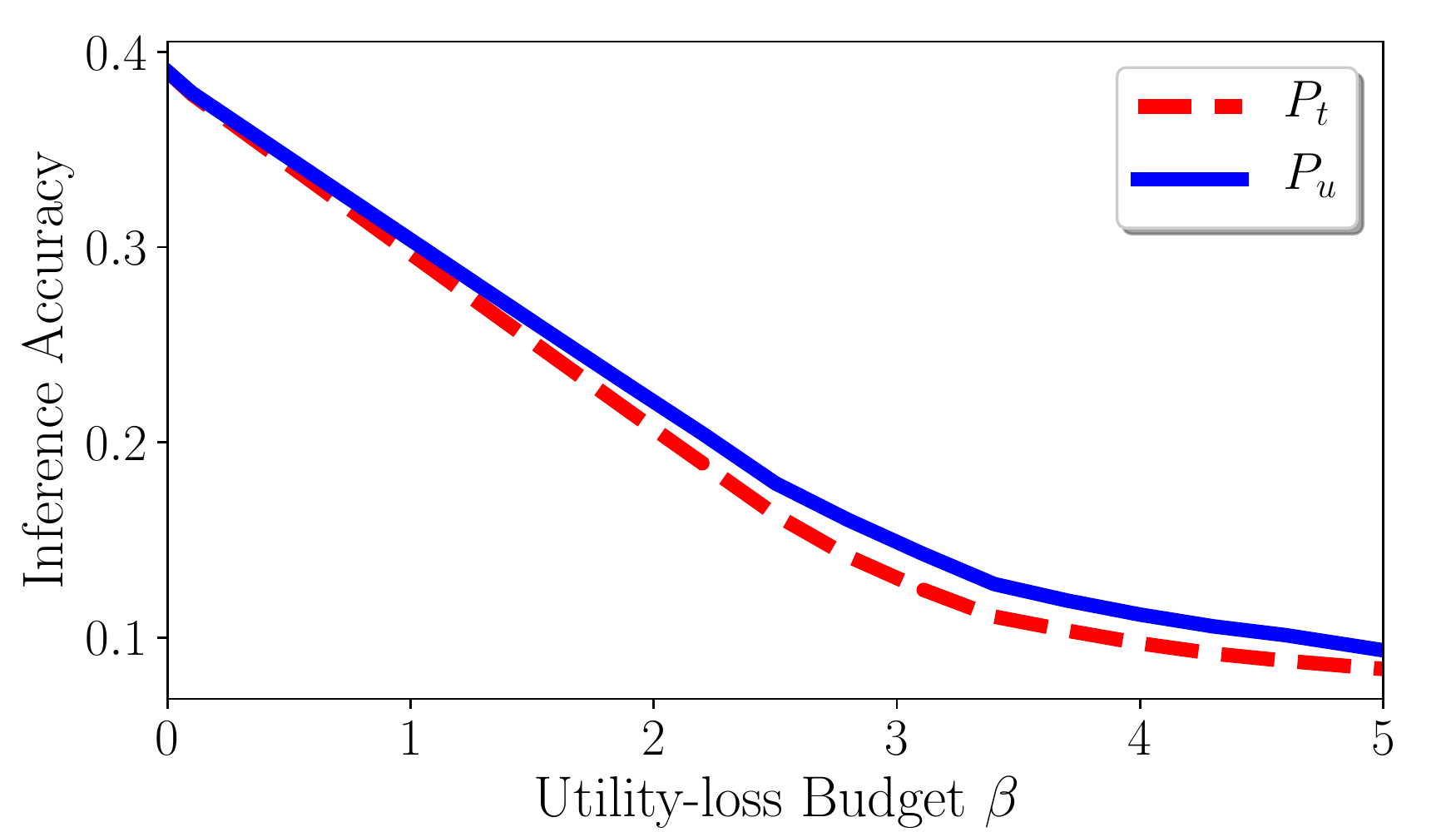}}
\caption{Impact of the target probability distribution. The attack is NN-A and the defender uses LR-D.} 
\label{tpd}
\end{figure}

\begin{figure}[t]
\centering
{\includegraphics[width=0.4 \textwidth]{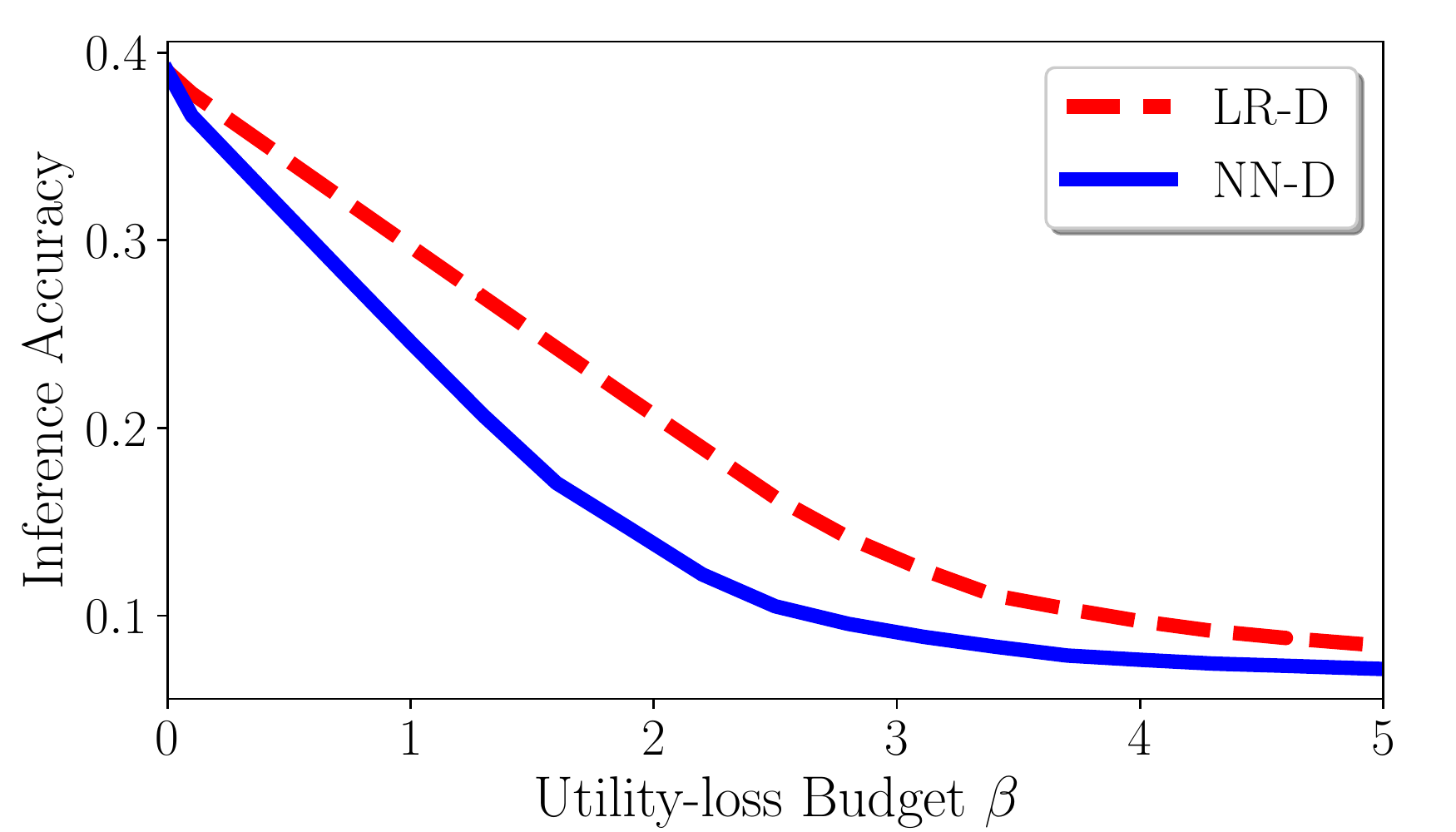}}
\caption{Impact of the defender's classifier. The attack is NN-A.} 
\label{dclassifier}
\end{figure}

\myparatight{Impact of the target probability distribution} Figure~\ref{tpd} compares the performance of the two target probability distributions. We observe that the target probability distribution $\mathbf{p}_t$ outperforms $\mathbf{p}_u$, especially when the utility-loss budget is relatively large. 
Specifically, the attacker's inference accuracy is smaller when the defender uses $\mathbf{p}_t$.  This is because $\mathbf{p}_t$ considers the attribute information in the training dataset, while $\mathbf{p}_u$ assumes no knowledge about attributes. Specifically, according to our solution in Equation~\ref{ex1}, the defender adds the noise $\mathbf{r}_i$ with a probability that is the corresponding target probability normalized by the magnitude of $\mathbf{r}_i$. Suppose the defender's classifier predicts city $j$ for a user, where $j$ is likely to be the true attribute value of the user since the defender's classifier is relatively accurate.  The noise $\mathbf{r}_j$ is 0. Roughly speaking, if the defender adds 0 noise, then the attacker is likely to infer the true attribute value. For the users whose true attribute values are rare (i.e., small fraction of users have these attribute values), the defender is less likely to add 0 noise when using $\mathbf{p}_t$ than using $\mathbf{p}_u$. As a result, the attacker has a lower inference accuracy when $\mathbf{p}_t$ is used.

\myparatight{Impact of the defender's classifier} Figure~\ref{dclassifier} shows the attacker's inference accuracy when the defender uses different classifiers, where the attack is NN-A. We observe that when the defender chooses the NN-D classifier, the attacker's inference accuracy is lower with the same utility-loss budget. One reason is that the noise found in Phase I is more likely to transfer between classifiers in the same category. Specifically, the noise optimized based on the neural network classifier NN-D is more likely to transfer to the neural network classifier NN-A than the logistic regression classifier LR-A. 

\begin{figure}[!t]
\centering
{\includegraphics[width=0.4 \textwidth]{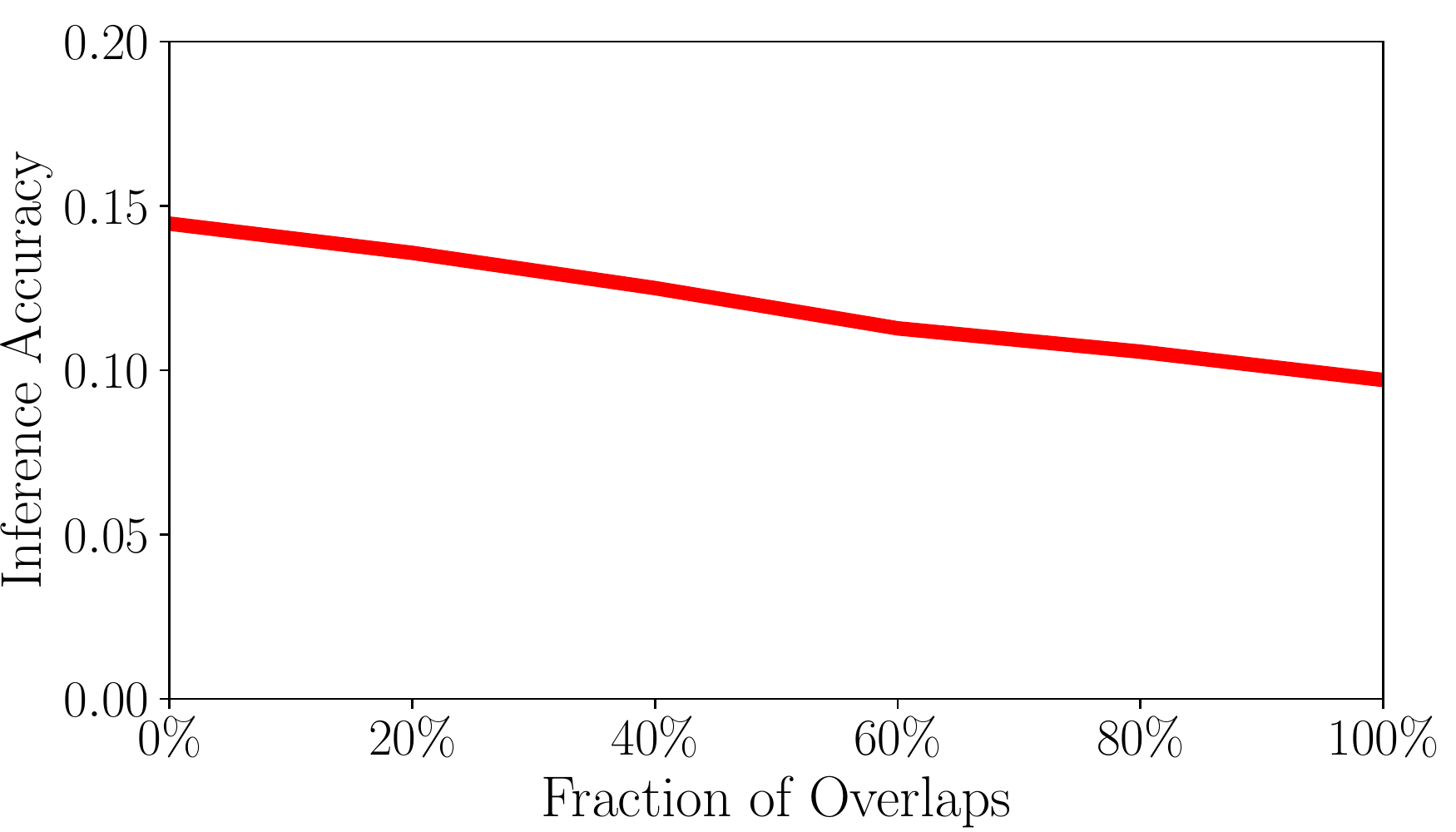}}
\caption{Impact of the overlap between the training datasets used by the attacker and the defender. }
\label{training}
\end{figure}

\begin{figure}[!t]
\centering
{\includegraphics[width=0.4 \textwidth]{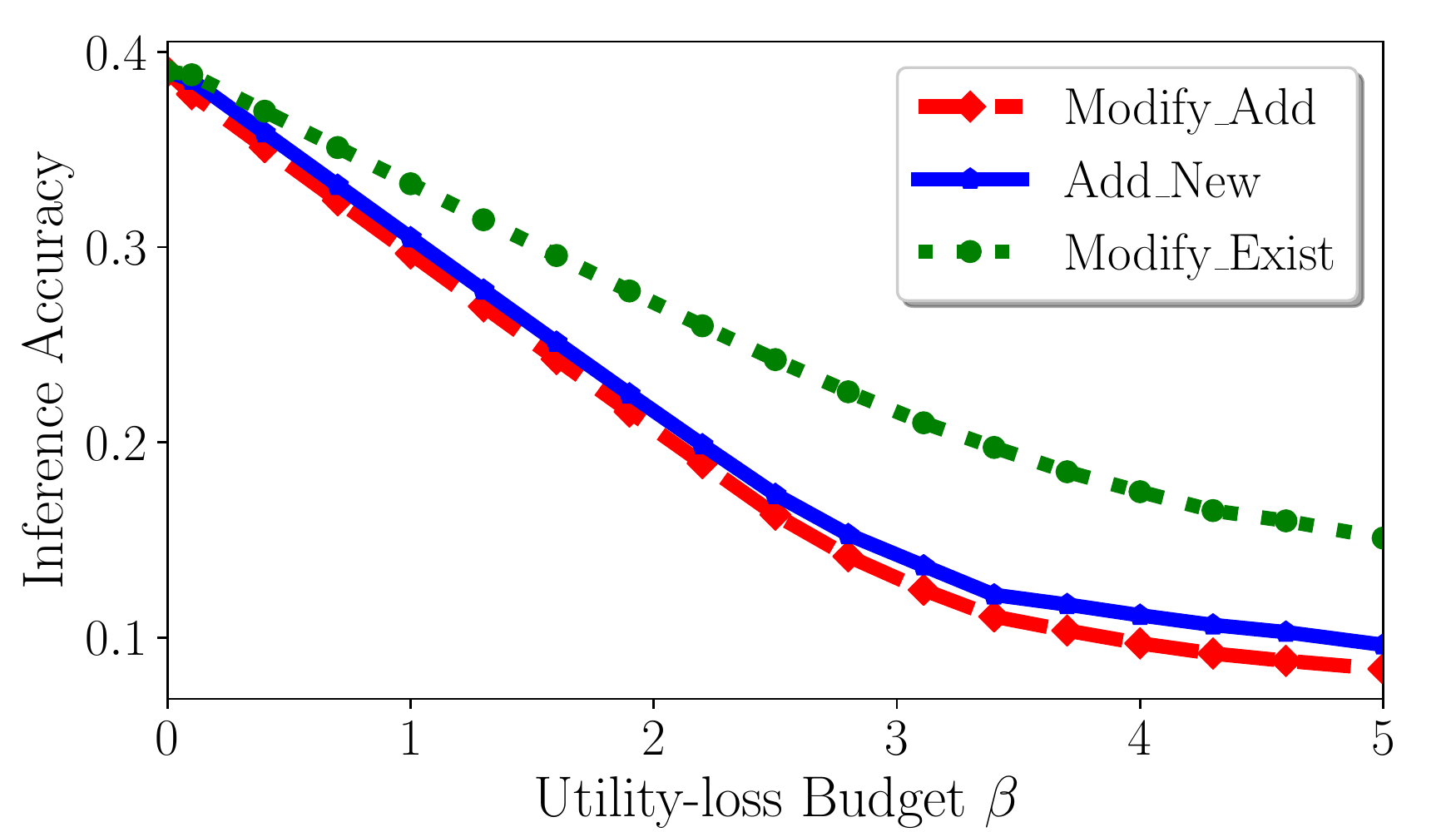}}
\caption{Impact of different noise-type-policies.} 
\label{policy}
\end{figure}

\myparatight{Impact of different training datasets} In practice, the attacker and the defender may use different training datasets to train their classifiers.  We randomly and evenly split the training dataset into two folds with $\alpha\%$ overlap, where $\alpha\%$ ranges from 0\% to 100\%. We consider the attack NN-A and use one fold to train the classifier, while we consider the defender's classifier is LR-D and use the other fold to train it. We set the utility-loss budget to be 4, which reduces most attacks to be close to the baseline attack. 
Figure~\ref{training} shows the attacker's inference accuracy as a function of the overlap $\alpha\%$. We find that the differences between the training datasets used by the attacker and the defender have impact on the effectiveness of AttriGuard, but the impact is small. Specifically, when the defender and the attacker use the same training dataset to learn their classifiers, the attacker's inference accuracy is around 0.10. The attacker's inference accuracy increases when the overlap between the training datasets decreases, but the attacker's inference accuracy is still less than 0.15 even if there are no overlaps. The reason is that both the attacker's classifier and the defender's classifier model the relationships between public data and attributes. Once both of their (different) training datasets are representative, the noise optimized based on the defender's classifier is very likely to transfer to the attacker's classifier.

\myparatight{Impact of different noise-type-policies} Figure~\ref{policy} compares the three noise-type-policies. Modify\_Add outperforms Add\_New, which outperforms Modify\_Exist. This is because Modify\_Add is the most flexible policy, allowing AttriGuard to modify existing rating scores or add new rating scores.  A user often reviews a very small fraction of apps (e.g., 0.23\% of apps on average in our dataset), so Add\_New is more flexible than Modify\_Exist, making Add\_New outperform Modify\_Exist.

\begin{figure}[!t]
\centering
{\includegraphics[width=0.4 \textwidth]{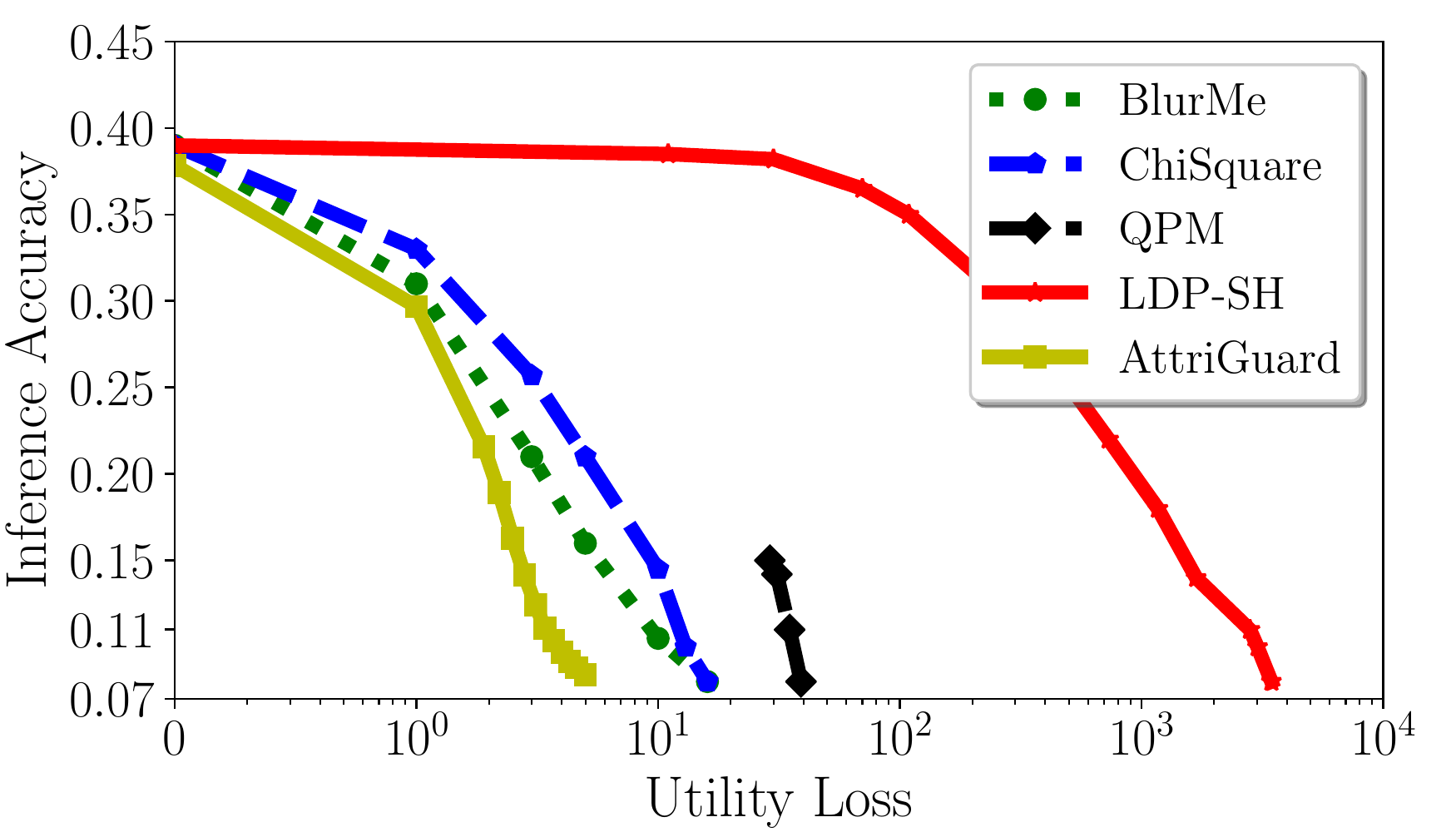}}
\caption{Comparing AttriGuard with existing defense methods.}
\vspace{-2mm}
\label{comparison}
\end{figure}

\myparatight{Comparing AttriGuard with existing defense methods} Figure~\ref{comparison} compares AttriGuard with existing defense methods developed by different research communities: BlurMe~\cite{weinsberg2012blurme}, ChiSquare~\cite{ChenObfuscationPETS14}, Quantization Probabilistic Mapping (QPM)~\cite{Salamatian:2015}, and Local Differential Privacy-Succinct Histogram (LDP-SH)~\cite{BassilySuccinctHistograms15}.  BlurMe and ChiSquare select the apps based on their correlations with the attribute values (i.e., cities in our case) that do not belong to the user and change the rating scores for the selected apps. QPM is an approximate solution to a game-theoretic formulation. We quantize public data to 200 clusters in QPM. 
LDP-SH is a local differential privacy method for categorical data. In our case, each entry of $\mathbf{x}$ can be viewed as categorical data taking values 0, 0.2, 0.4, 0.6, 0.8, or 1.0. We apply LDP-SH to each entry of $\mathbf{x}$.  We didn't use other LDP methods~\cite{Erlingsson:2014,Wang17LDP} because they do not preserve the semantics of rating scores. For instance, to obfuscate a user's rating score to an app, RAPPOR~\cite{Erlingsson:2014} might generate several rating scores for the app for the user, which is unrealistic. All the compared methods except LDP-SH use the same training dataset, while LDP-SH does not need training dataset. We note that BlurMe and ChiSquare require the defender to know users' true private attribute values.  


Each compared method has a parameter to control privacy-utility tradeoffs. 
For a method and a given parameter value, the method adds noise to users' public data, and we can obtain a pair (utility loss, inference accuracy), where the utility loss and inference accuracy are averaged over all test users. Therefore, for each method, 
via setting a list of different parameter values, we obtain a list of pairs (utility loss, inference accuracy). Then, we plot these pairs as a utility loss vs. inference accuracy curve. Figure~\ref{comparison} shows the curve for each method. 

Our AttriGuard outperforms all compared defense methods. Specifically, to achieve the same inference accuracy, AttriGuard adds substantially smaller noise to public data. AttriGuard outperforms BlurMe and ChiSquare because they add noise to entries of $\mathbf{x}$ that are selected based on heuristics, while AttriGuard adds minimum noise via solving optimization problems. 
We explored a large range of the parameter to control privacy-utility tradeoffs for QPM, but QPM cannot reach to the low utility-loss region, i.e., we only observe a short curve for QPM in Figure~\ref{comparison}. This is because quantization changes public data substantially, which is equivalent to adding large noise. AttriGuard outperforms LDP-SH because LDP-SH aims to achieve a privacy goal that is different from defending against attribute inference attacks.

\begin{figure}[!t]
\centering
{\includegraphics[width=0.4 \textwidth]{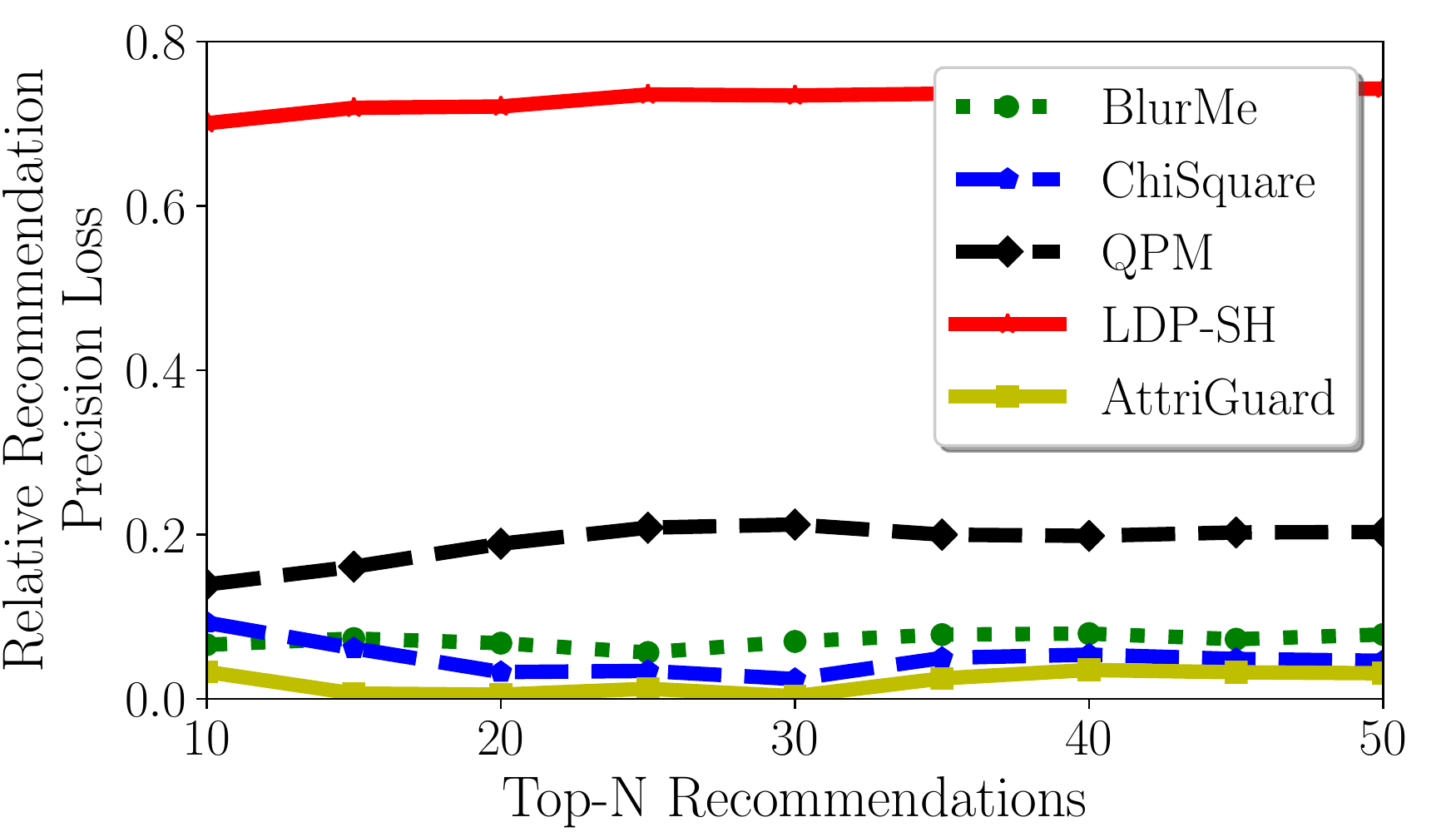}\label{recsys}}
\caption{Relative recommendation precision loss vs. top-$N$ recommendations.}
\vspace{-2mm}
\label{recsys}
\end{figure}

\myparatight{Utility loss for recommender systems} We evaluate the utility loss of the public data when they are used for recommender systems. 
For each user in the training and testing datasets, we randomly sample 5 of its rated apps to test a recommender system. We use a standard matrix factorization based recommender system to recommend top-$N$ items for each user. We implemented the recommender system using the code from http://surpriselib.com/. We measure the performance of the recommender system using a standard metric, i.e., \emph{recommendation precision}. For each user, the recommendation precision is the fraction of its recommended top-$N$ items that are among the sampled 5 rated apps. The recommendation precision for the entire recommender system is the recommendation precision averaged over all users.


For each compared defense method, we use the defense method to add noise to the testing users, where the noise level is selected such that an attacker's inference accuracy is close to 0.1 (using the results in Figure~\ref{comparison}). 
Then, for each compared defense method, we compute the \emph{relative recommendation precision loss} defined as $\frac{|Pre_1 - Pre_2|}{Pre_1}$, where $Pre_1$ and $Pre_2$ are the recommendation precisions before and after adding noise, respectively.
 Figure~\ref{recsys} shows the {relative recommendation precision loss} as a function of $N$ for the compared methods. We observe AttriGuard outperforms the compared methods. Moreover, our results indicate that $L_0$ norm of the noise is a reasonable utility-loss metric for recommender systems, as a method with larger $L_0$-norm noise also has larger relative recommendation precision loss. One exception is the comparison between BlueMe and ChiSquare: ChiSquare adds noise with larger $L_0$ norm but has lower relative recommendation precision loss. This means that ChiSquare adds noise that is more similar to a user's public data and thus has less impact on a user's profile of preferences.




\section{Discussions and Limitations}
\label{discussion}


\myparatight{Approximating the game-theoretic optimization problems} 
One natural direction is to find approximate solutions to the intractable game-theoretic optimization problems. Our experiments demonstrated that the existing approximate solution called QPM~\cite{Salamatian:2015} incurs larger utility loss than our AttriGuard. We note that we could apply the idea of AttriGuard to approximate the game-theoretic optimization problem in Equation~\ref{gametheoryformulation} in Appendix~\ref{gametheory}. However, such approximation is not meaningful. Specifically, AttriGuard essentially finds the noise mechanism for a given user (i.e., $\mathbf{x}$ is fixed) and treats the mechanism as a probability distribution over the representative noise. However, if we fix $\mathbf{x}$ and assume the probabilistic mapping to be a probability distribution over the representative noise in Equation~\ref{gametheoryformulation}, then the objective function in the optimization problem becomes a constant. In other words, any probabilistic mapping that satisfies the utility-loss budget is an approximate solution, which is not meaningful. 
We believe it is an interesting future work to study better approximate solutions to the game-theoretic optimization problems, e.g., the one in Equation~\ref{gametheoryformulation} in Appendix~\ref{gametheory}. 

\myparatight{Detecting noise} An attacker could first detect the noise added by AttriGuard and then perform attribute inference attacks. In our experiments, we tried  a low-rank approximation based method to detect noise and AttriGuard is still effective against the method. However, we acknowledge that this does not mean an attacker cannot perform better attacks via detecting the noise. We believe it is an interesting future work to systematically study the possibility of detecting noise both theoretically and empirically.   We note that detecting noise in our problem is different from \emph{detecting adversarial examples}~\cite{detection1,xu2017feature,MagNet,Madetection18,HeICLR18} in adversarial machine learning, because detecting adversarial examples is to detect whether a given example has attacker-added noise or not.  However, detecting adversarial examples may be able to help perform better attribute inference attacks. Specifically, if an attacker detects that a public data vector is an adversarial example, the attacker can use a defense-aware attribute inference attack for the public data vector, otherwise the attacker can use a defense-unaware attack.    

\myparatight{Interacting with adversarial machine learning} An attacker could use robust classifiers, which are harder to evade, to infer user attributes. In our experiments, we evaluated three  robust classifiers: adversarial training,  defensive distillation, and region-based classification. However, our defense is still effective for attacks using such robust classifiers. 
As the adversarial machine learning community develops more robust classifiers, an attacker could leverage them to infer attributes. However, we speculate that robust classifiers could always be evaded with large enough noise. In other words, we could still leverage evasion attacks to defend against attribute inference attacks, but we may need larger noise (thus larger utility loss) when the attacker uses a robust classifier that is harder to evade.

\myparatight{Multiple attributes} When users have multiple attributes, an attacker could leverage the correlations between attributes to perform better attribute inference attacks. The defender can design the target probability distribution based on the joint probability distribution of attributes to protect users against such attacks.

\myparatight{Dynamic public data} In this work, we focus on one-time release of the public data. It would be interesting to extend our framework to dynamic public data. For dynamic public data, an attacker could learn more information and perform better attribute inference attacks when observing historical public data.

\section{Conclusion and Future Work}
In this work, we propose a practical two-phase framework called \emph{AttriGuard} to defend against attribute inference attacks. In Phase I, AttriGuard finds a minimum noise for each attribute value via an evasion attack that we optimize to incorporate the unique characteristics of privacy protection. In Phase II, AttriGuard randomly selects one of the noise found in Phase I to mislead the attacker's inference. 
Our empirical results on a real-world dataset demonstrate that 1) we can defend against attribute inference attacks with a small utility loss, 2) adversarial machine learning can play an important role at privacy protection, and 3) our defense significantly outperforms existing defenses.


Interesting directions for future work include 1) studying the possibility of detecting the added noise both theoretically and empirically, 2) designing better approximate solutions to the game-theoretic optimization problems, and 3) generalizing AttriGuard to dynamic and non-relational public data, e.g., social graphs.  

\noindent
{\bf Acknowledgements:} We thank the anonymous reviewers for insightful reviews.

{
\balance{
\bibliographystyle{unsrt}
\bibliography{refs}
}}



\appendices

\section{Game-Theoretic Formulation}
\label{gametheory}
Shokri et al.~\cite{ShokriCCS12} proposed a game-theoretic formulation for defending against location inference attacks. In location inference attacks, both the public data and private attribute are users' true locations. Specifically, a user's true public data is the user's true location; the defender obfuscates the true location to a fake location; and the attacker aims to infer the user's true location, which can also be viewed as the user's private attribute. The game-theoretic formulation defends against the optimal location inference attack that adapts based on the knowledge of the defense. We extend this game-theoretic formulation for attribute inference attacks. In attribute inference attacks, public data and private attributes are different. 



\subsection{Notations} 
We denote by $s$ and $\mathbf{x}$ the private attribute and public data, respectively. 
We denote by $\text{Pr}(s,\mathbf{x})$ the joint probability distribution of  $s$ and $\mathbf{x}$. 
The defender aims to find a probabilistic mapping $f$, which obfuscates a true public data $\mathbf{x}$ to a noisy public data $\mathbf{x}^\prime$ with a probability $f(\mathbf{x}^\prime|\mathbf{x})$. The probabilistic mapping $f$ is essentially a matrix, whose number of rows and number of columns is the domain size of the public data vector $\mathbf{x}$. 

\subsection{Privacy Loss}
Suppose a user's true private attribute value is $s$ and an attacker infers the user's private attribute value to be $\hat{s}$. We denote the privacy loss for the user as a certain metric $d_p(s,\hat{s})$. 
For example, one choice for the privacy loss metric could be:
\begin{align}
d_p(s,\hat{s})=
\begin{cases}
1 &\text{ if } s=\hat{s} \\
0 &\text{ otherwise,}
\end{cases}
\end{align}
which means that the privacy loss is 1 if the attacker correctly infers the user's attribute value, and 0 otherwise.

\subsection{Utility Loss} 
For a true public data vector $\mathbf{x}$ and its corresponding noisy vector $\mathbf{x}^\prime$, we define the utility loss as $d_q(\mathbf{x},\mathbf{x}^\prime)$, which could be any distance metric over $\mathbf{x}$ and $\mathbf{x}^\prime$. For instance, $d_q(\mathbf{x},\mathbf{x}^\prime)$ could be the $L_0$ norm of the noise $||\mathbf{x}^\prime-\mathbf{x}||_0$, which is the number of entries of $\mathbf{x}$ that are modified. Given the marginal probability distribution $\text{Pr}(\mathbf{x})$ and the probabilistic mapping $f$, we have the expected utility loss as follows:
\begin{align}
L=\sum_{\mathbf{x},\mathbf{x}^\prime}\text{Pr}(\mathbf{x}) f(\mathbf{x}^\prime|\mathbf{x})d_q(\mathbf{x}^\prime,\mathbf{x}).
\end{align}

\subsection{Defender's Strategy}
The defender aims to construct a probabilistic mapping $f$ to defend against the optimal inference attack subject to a utility-loss budget $\beta$. 
%
%
%
The attacker knows the joint probability distribution $\text{Pr}(s,\mathbf{x})$ and the probabilistic mapping $f$. 
After observing a noisy public data vector $\mathbf{x}^\prime$, the attacker can compute a posterior probability distribution of the private attribute $s$ as follows:
\begin{align}
\text{Pr}(s|\mathbf{x}^\prime)&=\frac{\text{Pr}(s,\mathbf{x}^\prime)}{\text{Pr}(\mathbf{x}^\prime)} \\
&=\frac{\sum_{\mathbf{x}}\text{Pr}(s,\mathbf{x})f(\mathbf{x}^\prime|\mathbf{x})}{\text{Pr}(\mathbf{x}^\prime)}
\end{align}
Suppose the attacker infers the private attribute to be $\hat{s}$. Then, the conditional expected privacy loss is $\sum_{s}\text{Pr}(s|\mathbf{x}^\prime)d_p(s,\hat{s})$. 
Therefore, the maximum conditional expected privacy loss is as follows:
\begin{align}
\max_{\hat{s}}\sum_{s}\text{Pr}(s|\mathbf{x}^\prime)d_p(s,\hat{s})
\end{align}
Considering the probability distribution of $\mathbf{x}^\prime$, we have the unconditional expected privacy loss as follows:
\begin{align}
&\sum_{\mathbf{x}^\prime}\text{Pr}(\mathbf{x}^\prime)\max_{\hat{s}}\sum_{s}\text{Pr}(s|\mathbf{x}^\prime)d_p(s,\hat{s})\nonumber \\
=&\sum_{\mathbf{x}^\prime}\max_{\hat{s}}\sum_s\sum_{\mathbf{x}}\text{Pr}(s, \mathbf{x})f(\mathbf{x}^\prime|\mathbf{x})d_p(s,\hat{s}).
\end{align}
We define $y_{\mathbf{x}^\prime}=\max_{\hat{s}}\sum_s\sum_{\mathbf{x}}\text{Pr}(s, \mathbf{x})f(\mathbf{x}^\prime|\mathbf{x})d_p(s,\hat{s})$. The defender's goal is to minimize the unconditional expected privacy loss subject to a utility-loss budget. Formally, the defender aims to solve the following optimization problem:
\begin{align}
&\min \sum_{\mathbf{x}^\prime} y_{\mathbf{x}^\prime} \\
\text{subject to } &L \leq \beta.
\end{align}
According to Shokri et al.~\cite{ShokriCCS12}, this optimization problem can be transformed to the following linear programming problem: 
\begin{align}
&\min \sum_{\mathbf{x}^\prime} y_{\mathbf{x}^\prime} \nonumber \\
\text{subject to } &L \leq \beta \nonumber\\
& y_{\mathbf{x}^\prime} \geq \sum_s\sum_{\mathbf{x}}\text{Pr}(s, \mathbf{x})f(\mathbf{x}^\prime|\mathbf{x})d_p(s,\hat{s}), \forall \mathbf{x}^\prime, \hat{s} \nonumber\\
& \sum_{\mathbf{x}^\prime}f(\mathbf{x}^\prime|\mathbf{x}) = 1, \forall \mathbf{x} \nonumber\\
\label{gametheoryformulation}
& f(\mathbf{x}^\prime|\mathbf{x}) \geq 0, \forall \mathbf{x}, \mathbf{x}^\prime
\end{align}

\subsection{Limitations}
The formulated optimization problem is computationally intractable for attribute inference attacks in practice. Specifically, the computation cost is \emph{exponential} to the dimensionality of the public data vector, which is often high in practice. For instance, in recommender systems, a public data vector consists of a user's rating scores to the items that the user rated and 0 for the items that the user did not rate. Suppose a recommender system has 100 items (this is a very small recommender system in practice) and a rating score can be 1, 2, 3, 4, or 5. Then, the domain size of the public data vector $\mathbf{x}$ is $6^{100}$ and the size of the probabilistic mapping matrix $f$ is $6^{100} \times 6^{100}=6^{200}$. Therefore, even in the context of a very small recommender system with 100 items, it is intractable to solve the formulated optimization problem. 

\end{document}